\documentclass[11pt,a4paper]{article}
\pdfoutput=1

\usepackage{jheppub}
                     
\usepackage[mathscr]{euscript}
\usepackage[textsize=scriptsize,textwidth=2.5cm]{todonotes}

\usepackage{amssymb,amsmath,latexsym,bm,amsfonts}
\usepackage{graphicx}
\usepackage{longtable}
\usepackage{color,xcolor}
\usepackage{indentfirst}
\usepackage{subfig}
\usepackage{comment}
\usepackage{float}
\usepackage[normalem]{ulem}
\usepackage{array}

\usepackage{relsize}
\usepackage{graphicx}

\numberwithin{equation}{section}
\allowdisplaybreaks

\makeatletter
\def\@fpheader{\phantom{Prepared for submission to JHEP}}
\makeatother

\newcommand{\bea}{{\begin{eqnarray}}}
\newcommand{\eea}{{\end{eqnarray}}}
\newcommand{\mA}{{\cal A}}
\newcommand{\mH}{{\cal H}}
\newcommand{\mM}{{\cal M}}
\newcommand{\mL}{{\cal L}}
\newcommand{\mP}{{\cal P}}

\newcommand{\be}{\begin{equation}}
\newcommand{\ee}{\end{equation}}
\newcommand{\bpm}{\begin{pmatrix}}
\newcommand{\epm}{\end{pmatrix}}

\newcommand{\EV}[1]{\langle #1 \rangle}
\newcommand{\beqn}{\begin{eqnarray}}
\newcommand{\eeqn}{\end{eqnarray}}

\newcommand{\p}{\partial}

\newcommand{\ba}{\begin{aligned}}
\newcommand{\ea}{\end{aligned}}
\newcommand{\bi}{\begin{enumerate}}
\newcommand{\ei}{\end{enumerate}}

\def\Bra#1{\left\langle#1\right|}
\def\Ket#1{\left|#1\right\rangle}
\def\Vev#1{\left\langle#1\right\rangle}
\newcommand{\eq}[1]{\begin{align}#1\end{align}}
\newcommand{\eqsp}[1]{\begin{equation}\begin{split}#1\end{split}\end{equation}}
\newcommand{\eqga}[1]{\begin{equation}\begin{gathered}#1\end{gathered}\end{equation}}

\newcommand{\zc}{z}

\DeclareMathOperator{\Tr}{Tr}
\DeclareMathOperator{\tr}{tr}

\title{Modular Hamiltonians in flat holography and (W)AdS/WCFT} 
\author[a]{Luis Apolo,} 
\author[b]{Hongliang Jiang,} 
\author[a,c]{Wei Song,} 
\author[a]{and Yuan Zhong} 

\affiliation[a]{Yau Mathematical Sciences Center, Tsinghua University, Beijing 100084, China}
 
\affiliation[b]{Albert Einstein Center for Fundamental Physics, Institute for Theoretical Physics, University of Bern, Sidlerstrasse 5, 3012 Bern, Switzerland}
 
\affiliation[c]{Institute for Advanced Study, 1 Einstein Drive, Princeton, NJ 08540, USA}
  \emailAdd{apolo@mail.tsinghua.edu.cn, jiang@itp.unibe.ch, wsong2014@mail.tsinghua.edu.cn, zhongy17@mails.tsinghua.edu.cn} 
 
\abstract{
We study several aspects of holographic entanglement in two models known as flat$_3$/BMSFT and (W)AdS$_3$/WCFT. These are two examples of holography beyond AdS/CFT where the boundary field theories are not Lorentz invariant but still feature an infinite set of local symmetries. In the first example, BMS-invariant field theories (BMSFTs) are conjectured to provide a holographic description of quantum gravity in asymptotically flat three-dimensional spacetimes; while in the second example, warped conformal field theories (WCFTs) are proposed to describe quantum gravity in warped AdS$_3$ or AdS$_3$ backgrounds with Dirichlet-Neumann boundary conditions. In particular, we derive the modular Hamiltonian for single intervals in both BMSFTs and WCFTs and find the holographic duals in the bulk using the covariant formulation of gravitational charges. We also extend the first law of entanglement entropy to these models of non-AdS holography and discuss the bound on ``modular chaos'' introduced recently in the context of the AdS/CFT correspondence.
}

\begin{document}
\maketitle
\flushbottom

\section{Introduction} \label{se:introduction}

The modular Hamiltonian plays an important role in the study of entanglement entropy in both quantum field theory and holography. It can be formally defined as 
\eq{
{\cal H}_{mod} = -\log \rho_{\mA},
} 
where $\rho_{\mA}$ is the reduced density matrix of a state on a subregion $\mA$. The modular Hamiltonian generates a symmetry transformation between operators in the causal domain of dependence of $\mathcal{A}$. In general, this symmetry is not geometrically realized and the modular Hamiltonian is not a local operator. Nevertheless,  there are several similarities between modular Hamiltonians and ordinary Hamiltonians. For example, up to a normalization, the R\'{e}nyi entropy $S^{(n)}_{\cal A}=(1-n)^{-1}\log \tr \big(e^{-n {\cal H}_{mod}}\big)$ resembles the free energy $\mathcal F (T)$ of a thermodynamic system with Hamiltonian $\mathcal H_{mod}$ and temperature $T = 1/n$. Similarly, the entanglement entropy resembles the thermal entropy and satisfies the thermodynamical relation $S = -\p \mathcal F/ \p T$,
\eq{
S_{\mA} &= -\tr \big( \rho_{\cal A}\ln\rho_{\cal A} \big) =  \lim_{n\to1} \p_{1/n} \Big[ \frac{1}{n} \log \tr \big(e^{-n {\cal H}_{mod}} \big)\Big].
} 
In some special cases including the Rindler wedge in QFTs~\cite{Bisognano:1975ih,Bisognano:1976za} and ball-shaped regions in CFTs~\cite{Casini:2011kv}, the modular Hamiltonian on the vacuum state becomes an ordinary Hamiltonian that generates a geometrical flow and can be written as a spacetime integral of local operators. 

The modular Hamiltonian has been useful in several aspects of the AdS/CFT correspondence, including the derivation of holographic entanglement entropy~\cite{Casini:2011kv}, verifying the first law and various entropy inequalities~\cite{Blanco:2013joa,Lashkari:2016idm,Blanco:2017akw}, deriving the gravitational equations of motion from entanglement~\cite{Lashkari:2013koa,Faulkner:2013ica, Faulkner:2017tkh,Jiang:2019qvd}, proving the averaged and quantum null energy conditions~\cite{Faulkner:2016mzt,  Koeller:2017njr,Balakrishnan:2017bjg,Ceyhan:2018zfg}, and generating explicit prescriptions for bulk reconstruction~\cite{Jafferis:2015del, Faulkner:2017vdd, Dong:2016eik, Harlow:2016vwg, Faulkner:2018faa}. Considering the significant role of the modular Hamiltonian in the AdS/CFT correspondence, we would like to extend this and related concepts to models of holography beyond AdS/CFT.
 
One of the goals of this paper is to provide a general prescription for determining the vacuum modular flow generator and the corresponding modular Hamiltonian in a specific class of models of non-AdS holography. On the field theory side, we will show that the modular flow generator $\zeta$ can be determined by requiring that ($i$) $\zeta$ is a linear combination of the vacuum symmetry generators; ($ii$) the diffeomorphism generated by $\zeta$ leaves the causal domain $\mathcal D$ of the interval $\mathcal A$ invariant; and ($iii$) the flow $e^{i\zeta}$ maps each point in $\mathcal D$ back to itself. The modular Hamiltonian is then given by ${\cal H}_{mod} = {\cal H}_{\zeta} + const$, where ${\cal H}_{\zeta}$ is the Noether charge associated with the vector $\zeta$ while the constant term guarantees that $\mathrm{tr} \rho_{\cal A} = 1$. In particular, this prescription allows us to investigate the bound on modular chaos~\cite{deBoer:2019uem} and the first law of entanglement entropy in QFTs that are not Lorentz invariant. Using the holographic dictionary, we can extend the boundary modular flow generator into a Killing vector in the bulk and compute the modular Hamiltonian using the corresponding gravitational charge. This enables us to discuss the compatibility between the geometric picture for holographic entanglement entropy, the first law of entanglement, and the emergence of the gravitational equations of motion in the bulk.

We will apply the prescription described above to two models of holography beyond AdS/CFT dubbed flat$_3$/BMSFT and (W)AdS$_3$/WCFT. The flat$_3$/BMSFT correspondence conjectures that gravity in asymptotically flat three-dimensional spacetimes is dual to a BMS-invariant field theory (BMSFT) at the null boundary~\cite{Bagchi:2010eg,Bagchi:2012cy}. On the other hand, (W)AdS$_3$/WCFT conjectures that gravity in warped AdS$_3$ or AdS$_3$ spacetimes  with Dirichlet-Neumann boundary conditions is dual to a warped CFT --- a two-dimensional field theory invariant under a spacetime Virasoro-Kac-Moody symmetry~\cite{Hofman:2011zj,Detournay:2012pc}. Both of the aforementioned models share several features that depart from the AdS/CFT correspondence in interesting ways. For example, neither BMSFTs nor WCFTs are Lorentz invariant field theories and the causal domain of dependence of an interval $\mA$ consists of an infinitely long strip. As a result, the flow generated by the modular Hamiltonian differs significantly from the modular flow in AdS/CFT, both in the boundary and the bulk sides of these correspondences, as discussed in more detail later. Relatedly, the bulk dual of entanglement entropy in these models consists of a spacelike geodesic --- the set of fixed points of the bulk modular flow --- that hangs from the boundary interval $\mA$ by a set of null geodesics parallel to the modular flow~\cite{Song:2016gtd,Jiang:2017ecm}. A general holographic entropy proposal based on these so-called \emph{swing surfaces} is given in~\cite{Apolo:2020bld}.

In this paper we will derive the modular Hamiltonian on the vacuum and thermal states for a single interval in flat$_3$/BMSFT and (W)AdS$_3$/ WCFT. The explicit expressions for the modular Hamiltonian allow us to gain further insights into entanglement in both the bulk and boundary sides of these models. For instance, a bound on modular chaos for relativistic QFTs has been recently proposed in~\cite{deBoer:2019uem}. We will show that this bound is satisfied in both BMSFTs and WCFTs despite the fact that these theories are not Lorentz invariant and the existence of such a bound may not be a priori expected. We will also use the modular Hamiltonian to derive the first law of entanglement entropy for highly-symmetric states in WCFTs directly from a field theory calculation. On the bulk side we will show how the modular flow generator of the vacuum and thermal states can be extended into the bulk and describe how it can be used to determine the swing surface whose area captures the entanglement entropy of an interval at the boundary. Finally, we compute the gravitational charge associated with the bulk modular flow generator and show that it matches the modular Hamiltonian at the boundary.  This allows us to provide a holographic derivation of the first law of entanglement entropy for special states in both flat$_3$/BMSFT and (W)AdS$_3$/WCFT.\footnote{The first law of entanglement entropy in flat$_3$/BMSFT was also discussed in \cite{Godet:2019wje, Fareghbal:2019czx}.}

The paper is organized as follows. In Section~\ref{se:modHnonAdS} we provide a general prescription for the derivation of the modular Hamiltonian in models of non-AdS holography, describe the swing surfaces whose areas reproduce the entanglement entropy of the dual field theories at the boundary, and discuss the holographic derivation of the first law. In Sections~\ref{se:bmsft} and~\ref{se:wcft} we apply these methods to the bulk and boundary sides of the flat$_3$/BMSFT and (W)AdS$_3$/WCFT correspondences, compute the modular Hamiltonian, derive the first law of entanglement entropy, and investigate the bound on modular chaos. 


\section{The modular Hamiltonian in non-AdS holography} \label{se:modHnonAdS}

In this section we review the generalized Rindler method proposed in~\cite{Jiang:2017ecm} and provide a general prescription for the calculation of the modular Hamiltonian associated with an interval $\mA$ at the boundary. We then extend the modular flow generator into the bulk and use it to determine the holographic entanglement entropy, which is given by the area of a swing surface that is homologous to $\mA$. We also propose that the gravitational charge associated with the bulk modular flow generator is holographically dual to the boundary modular Hamiltonian. Finally, using the holographic dictionary, we derive the first law of entanglement entropy from the bulk equations of motion and discuss the derivation of the equations of motion from entanglement.

\subsection{The boundary modular Hamiltonian} \label{se:modularH}
We begin by describing the generalization of the Rindler method~\cite{Casini:2011kv} proposed in~\cite{Castro:2015csg,Song:2016gtd,Jiang:2017ecm}. The latter uses the symmetries of the vacuum state to map an interval (and its causal domain) to a noncompact space whose generator of time translations is identified with the modular Hamiltonian. We also describe a general prescription to determine the modular flow generator that similarly exploits the symmetries of the vacuum. 

Let us consider a holographic correspondence between a gravitational theory on a $(d+1)$-dimensional manifold $\mM$ and a quantum field theory on its $d$-dimensional boundary $\p \mM$. The field theory is not necessarily local but it is invariant under some symmetry group $\sf G$. We assume that the vacuum state of the field theory exists and that it is invariant under a subgroup $\sf H \subset \sf G$. According to the holographic dictionary, $\sf H$ is the isometry group of the spacetime in the bulk that is dual to the vacuum state in the boundary, while $\sf G$ is the asymptotic symmetry group of the gravitational theory satisfying appropriate boundary conditions. Let $\mA$ denote a subregion of a co-dimension one boundary Cauchy surface $\p\Sigma \subset \p \mM$. The boundary of $\mA$ is denoted by $\p \mA$ and, in Lorentzian signature, its causal domain is denoted by $\mathcal{D} \subset \p\mM$.  

We further assume that a Hilbert space can be defined on a hypersurface $\p\Sigma \subset \p \mM$ such that, given a state $|\psi\rangle$ in $\p\Sigma$, we can define the reduced density matrix associated with $\mA$ by tracing out the degrees of freedom on its complement $\bar{\mA}$, namely
  \eq{
  \rho_\mA =\tr_{\bar\mA} |\psi\rangle \langle \psi|.
  }
Formally, we can define the modular Hamiltonian $\mH_{mod}$ by $\rho_{\mA}=e^{-\mH_{mod}}$ such that the entanglement entropy is given in terms of the reduced density matrix $\rho_{\mA}$ by 
	 \be S_{\cal A} = -\tr \big( \rho_{\cal A}\ln\rho_{\cal A} \big)=\tr \big(\rho_{\cal A}\mH_{mod} \big ).
	  \ee
In general, the modular Hamiltonian is nonlocal and the entanglement entropy is difficult to calculate. Nevertheless, in some cases that include ball-shaped regions on the vacuum state of CFTs, the modular Hamiltonian becomes a local operator~\cite{Casini:2011kv}. In these examples, the causal domain $\cal D$ can be mapped to a Rindler spacetime by a symmetry transformation. The entanglement entropy of the interval $\mA$ is then mapped to the thermal entropy of the Rindler space where  the modular Hamiltonian is the generator of translations along the Rindler timelike coordinate. This idea was generalized to nonrelativisitic quantum field theories with warped conformal symmetries in~\cite{Castro:2015csg,Song:2016gtd} and BMS symmetries in~\cite{Jiang:2017ecm}.

A generalized Rindler transformation $x\to \tilde{x}$ is a symmetry transformation in the field theory that satisfies the following properties~\cite{Jiang:2017ecm}:  
\begin{itemize}
\item[(\emph{i})] The transformation $x\to \tilde{x} = f(x)$ is a symmetry of the field theory where the domain of $f(x)$ is the causal domain $\mathcal{D}$ and its image is a noncompact manifold $\p\tilde{\mM}$, the generalized Rindler spacetime. Here $x$ and $\tilde{x}$ collectively denote the coordinates on $\p \mM$ and $\p \tilde \mM$, respectively.
\item[(\emph{ii})]  The transformation $x\to \tilde{x}$ is invariant under a pure imaginary (thermal) identification $(\tilde {x}^1,\dots,\tilde {x}^d)\sim (\tilde {x}^1+i\tilde{\beta}^1,\dots,\tilde {x}^d+i\tilde{\beta}^d)$. In particular, the vacuum is invariant under translations along the $\tilde{x}^i$ coordinates for which $\tilde{\beta}^i\neq0$, which guarantees that the vacuum state in $\cal{D}$ is mapped to a thermal state in $\p\tilde{\mM}$. Consequently, the generator of the thermal identification $\zeta \equiv \sum_i \tilde{\beta}^i\p_{\tilde{x}^i}$ leaves the vacuum invariant such that $\zeta =\sum_i a_i h_i$ where $a_i$ are constants and $h_i$ are the generators of $\sf H$.
\item[(\emph{iii})] The generator of the thermal identification $\zeta$ gives rise to a one-parameter flow $\tilde{x}^i[s]$ through the exponential map $e^{s\zeta}$. When $s$ is real, $\zeta$ generates a flow among points in $\p\mM$ that leaves the causal domain $\mathcal D$ and its boundary $\p\mathcal{D}$ invariant. On the other hand, a modular flow with $s=i/2$ maps a point in the causal domain $\cal D$ of $\mA$ to a point in the causal domain $\bar{\cal D}$ of the complement $\bar\mA$. Finally, the thermal identification $\tilde{x}^i[0]\sim \tilde{x}^i[i]$ maps any point back to itself and corresponds to $s=i$. 
\end{itemize}

If a generalized Rindler transformation satisfying the above properties exists, then we can define a Rindler time 
\eq{
\tau= \frac{1}{k} \sum_{i=1}^k 2\pi  (\tilde{\beta}^{i})^{-1} {\tilde x}^i, \qquad \tau \sim \tau + 2\pi i,\label{taubdy}
}
 where $k \le d$ is the total number of nonvanishing $\tilde{\beta}^i$ parameters. The partition function can then be written as $Z = \Tr e^{-2\pi {\cal H}_{\p_{\tau}}}$ where ${\cal H}_{\p_{\tau}}$ is the Noether charge generating translations along $\tau$. As a result, the modular Hamiltonian is geometrically realized and identified with the normalized Rindler time translation
\eq{
\zeta = 2\pi \p_\tau=\sum_{i} a_i h_i. \label{s2:zeta}
}
If we let ${\cal H}_i$ denote the Noether charges associated with the generators $h_i$ of $\sf H$, then the modular Hamiltonian is given in terms of the Noether charges  by ${\cal H}_{mod}= {\cal H}_{\zeta}+const $ where the constant term guarantees that $\Tr \rho_{\mA} = 1$ and
\eq{
{\cal H}_{\zeta} = \sum_{i} a_i \mathcal{H}_i. 
} 
Note that the inverse temperatures $\tilde{\beta}^i$ in the generalized Rindler transformation are arbitrary parameters that do not enter the final expressions for the modular Hamiltonian or the entanglement entropy. This is related to the fact that the modular flow generator $\zeta$ can be obtained without explicitly finding the Rindler transformation that maps the interval $\mA$ to the noncompact space $\p\tilde{\mM}$, as described in detail next.


\bigskip\noindent{\bf A general recipe for the modular flow generator.} We now provide a general prescription to determine the modular flow generator for the vacuum state without explicitly using the generalized Rindler map. Once we have an expression for the modular flow generator $\zeta$, it is straightforward to write down the corresponding modular Hamiltonian ${\cal H}_{mod}$. Given a subregion $\mA$ and its causal domain ${\cal D}$ in the field theory, we can determine the modular flow generator $\zeta$ by requiring it to satisfy the following properties:
\begin{itemize}
\label{steps}\item[(\emph{i})] The modular flow generator is a linear combination of the vacuum symmetry generators $h_i$ such that $\zeta=\sum_i a_ih_i$ where $a_i$ are arbitrary constants.
\item[(\emph{ii})] Up to an overall normalization, the $a_i$ coefficients are constrained by requiring that the modular flow generator leaves the boundary of the causal domain  $\p\mathcal{D}$ invariant. In particular, under the action of $\zeta$, an endpoint $p\in \p\mA$  can either be a fixed point of $\zeta$ or be moved along $\p\mathcal{D}$. 
\item[(\emph{iii})]\label{requirementiii} We can determine the overall normalization of $\zeta$ by requiring that $e^{i \zeta/2}$ maps a point in the interval $\mA$ (but not in $\p\cal \mA$) to its complement $\bar\mA$. Alternatively, we require that $e^{i \zeta }$ maps any point back to itself as this is equivalent to the periodicity condition on local Rindler time $\tau\sim\tau+2\pi i$.
\end{itemize} 
We will show that these properties determine the modular flow generator in BMSFTs and WCFTs in Sections~\ref{se:bmsft} and~\ref{se:wcft}, respectively, and that the results agree with known expressions obtained from the generalized Rindler map~\cite{Song:2016gtd,Jiang:2017ecm}. We will then use the modular flow generator to compute the corresponding modular Hamiltonians in these theories.


\subsection{A bound on modular chaos} \label{se:modularalgebra}

Let us now discuss perturbations of the modular flow generator $\zeta$ and their relationship to the bound on modular chaos~\cite{deBoer:2019uem}. It has been recently shown that in unitary and Lorentz-invariant theories, perturbations of the modular Hamiltonian $\delta {\cal H}_{mod}$ obtained from deformations of the state or the shape of the interval $\mA$ are exponentially bounded. More concretely, if we denote the matrix elements of $\delta {\cal H}_{mod}$ between any two states $\Ket{\psi_i}$ and $\Ket{\psi_j}$ on subregion $\mA$ by
  \eq{
  F_{ij}(s) = \big|\! \Bra{\psi_i} e^{i {\cal H}_{mod} s} \delta {\cal H}_{mod} e^{-i {\cal H}_{mod}s} \Ket{\psi_j} \!\big|,
  }
then the bound on modular chaos reads
  \eq{
  \lim_{s \to \pm \infty} \Big| \frac{d}{ds} \log F_{ij}(s) \Big| \le 2\pi.  \label{s2:bound}
  }
 A stronger version of this bound has also been conjectured to exist for large $N$ QFTs as shown in detail in~\cite{deBoer:2019uem}. The bound~\eqref{s2:bound} is saturated for deformations of the modular Hamiltonian satisfying the following kind of eigenvalue problem
 \eq{
 \big [{\cal H}_{mod}, \delta_\pm {\cal H}_{mod} \big] = \pm 2\pi  i   \delta_\pm {\cal H}_{mod}. \label{s2:modularchaos}
 }
These deformations are dubbed \emph{modular scrambling modes} in \cite{deBoer:2019uem}. The modular Hamiltonian of 2d CFTs has been shown to satisfy the bound on modular chaos in~\cite{Czech:2019vih,deBoer:2019uem}. Moreover, for ball shaped regions  in two and higher dimensional CFTs, perturbations of the modular Hamiltonian arising from shape deformations of $\mA$ along null directions have also been shown to saturate the bound on modular chaos~\cite{Casini:2017roe}.

As argued in the previous section, the modular Hamiltonian associated with the vacuum state is the Noether charge associated with the Killing vector $\zeta$. Let us now consider shape deformations of the entangling surface $\mA$ generated by certain linear combinations of the global symmetry generators that leave the vacuum invariant.  In this case, the eigenvalue problem~\eqref{s2:modularchaos} reduces to the problem of finding the set of eigenvectors of the modular flow generator $\zeta$ satisfying
\eq{
 \big [\zeta, \delta_\pm \zeta \big]= \pm 2\pi \,  \delta_\pm {\zeta},\label{s2:modularchaos2}
}
where $\delta _\pm  \zeta$ is a linear combination of the global symmetry generators $h_i$ with eigenvalue $\pm2\pi$, respectively. In Sections~\ref{se:bmsft} and~\ref{se:wcft}  we will show that variations of the modular Hamiltonian exist in both BMSFTs and WCFTs that saturate the bound on modular chaos and correspond to modular scrambling modes. This indicates that the bound on modular chaos  is also obeyed, at least for shape deformations on the vacuum state, in theories that violate Lorentz invariance, mirroring similar observations on the bound on chaos in WCFTs~\cite{Apolo:2018oqv}. We will also provide a bulk interpretation of the modular scrambling modes based on the geometric description of entanglement entropy described in the following section.


\subsection{Bulk modular flow and swing surfaces} \label{se:genrt}

In this section we extend the modular flow generator into the bulk and describe how it can be used to determine the bulk dual of the entanglement entropy of the boundary subregion $\cal A$. We begin by recalling that the modular flow generator~\eqref{s2:zeta} is a linear combination of the global symmetry generators $h_i \in \sf H$ that leave the vacuum of the field theory invariant. In the bulk, $\sf H$ is the isometry group of the spacetime dual to the vacuum.\footnote{One exception to this assumption is warped AdS$_3$ with Dirichlet boundary conditions~\cite{Compere:2014bia}. Although the asymptotic symmetry algebra in a restricted phase space is described by two Virasoro algebras, there is no bulk geometry with $SL(2,R)\times SL(2,R)$ isometries. Hence the discussion in this paper does not apply to this particular example. Nevertheless, a generalized gravitational entropy in this setup is discussed in ref.~\cite{Song:2016pwx} and the holographic entanglement entropy is also found to differ from the RT/HRT proposal.}  This means that each generator $h_i$ corresponds to an exact Killing vector $H_i$ in the bulk, such that the $H_i$ and $h_i$ generators satisfy the same symmetry algebra and 
  \eq{
  H_i \big|_{\p\mM} = h_i, \label{projection}
  }
where $H_i \big|_{\p\mM}$ denotes the vector $H_i$ at a cutoff surface $\p\mM$. More precisely, if we consider a cutoff surface in the bulk at $r = 1/\epsilon$ that is parametrized by coordinates $x^a$ with $a = 1, \dots, d$, then \eqref{projection} means that $H_i^a \big|_{r = 1/\epsilon} = h_i^a$.  As a result, the modular flow generator in the bulk corresponds to a Killing vector $\xi$ that is given by
\eq{
\xi =\sum_{i} a_i H_i ,\label{s2:xi}
}
where the $a_i$ coefficients are the same parameters featured in \eqref{s2:zeta} such that $\xi \big|_{\p\mM} = \zeta$. 

The boundary modular flow generator defines a boundary Rindler time \eqref{taubdy} with pure imaginary identifications which can be extended into the bulk by 
\be
\xi\equiv 2\pi \p_{\tau},\qquad \tau \sim \tau+2\pi i,
\ee
where $\tau$ now denotes the bulk Rindler time. Being a Killing vector field which implements purely imaginary identifications implies that the modular flow generator~\eqref{s2:xi} has a bifucating Killing horizon whose surface gravity is $2\pi$. The bifurcating surface is denoted by $\gamma_\xi$ such that $\xi|_{\gamma_\xi}=0$. The future/past directed light-sheets of $\gamma_\xi$ are denoted by $N_\pm$ and are defined as the null geodesic congruences with non-positive expansion that extend from $\gamma_\xi$ to the boundary ($\gamma_\xi$ has zero expansion as is therefore an extremal surface). From these assumptions and definitions we learn that:
\begin{itemize}
\item The light-sheets $N_\pm$ are half of the future/past Killing horizon and hence invariant under the flow generated by $\xi$. Using the fact that $\xi\big|_{\p\mM} = \zeta$ at the boundary, we find that the intersection of the light-sheets $N_+\cup N_-$ with ${\p \mM}$ corresponds to the boundary of the causal domain, namely $(N_+\cup N_-)\cap \p M=\p\mathcal{D}$. 
\item From its definition, $\xi$ is normal to the light-sheets $N_+\cup N_-$. As a result, we have   
\eq{ 
\xi_{[\mu} \nabla_\nu \xi_{\lambda]}\big|_{N_\pm}=0.  \label{normaleq1}
}
This is just Frobenius' theorem which  guarantees that the vector field is hypersurface orthogonal, see e.g.~Appendix B of~\cite{Wald:1984rg}.
\item 
From general properties of Killing horizons, the surface gravity on $N_\pm$ is a constant whose value depends on the normalization of the Killing vector $\xi$. The latter is fixed by the normalization of the modular flow generator at the boundary, namely $\xi|_{\p \mathcal M}=\zeta=2\pi \p_\tau$ with $\tau\sim \tau+2\pi i$. This means that the local Rindler time $\xi=2\pi \p_\tau$ in the bulk should also satisfy $\tau\sim \tau+2\pi i$. In other words, the surface gravity must be $2\pi$ such that
 \eq{
    \xi^\nu \nabla_\nu \xi^\mu \big|_{N_\pm} =\pm 2\pi \xi^\mu. \label{normaleq2}
  }
\item Finally, on the extremal surface $\gamma_{\xi}=N_+\cap N_-$, the modular flow generator satisfies
\be
  \xi \big |_{\gamma_\xi} = 0 , \qquad \nabla^{\mu}\xi^{\nu}\big|_{\gamma_\xi} = 2\pi n^{\mu\nu}, \label{surfacegravity}
\ee
where $n^{\mu\nu}=n_1^\mu n_2^\nu-n_2^\mu n_1^\nu$ is the binormal unit vector to $\gamma_\xi$. This follows from the fact that $\gamma_\xi$ is an extremal surface and from the normalization of $\xi$ in \eqref{normaleq2}, the latter of which has been determined by the normalization of $\zeta$ at the boundary.
\end{itemize}
We will verify that these properties of the bulk modular flow generator are satisfied in both flat$_3$/BMSFT and (W)AdS$_3$/WCFT in Sections~\ref{se:bmsft} and~\ref{se:wcft}, respectively.

\begin{figure}[h] 
  \centering
    \includegraphics[scale=0.55]{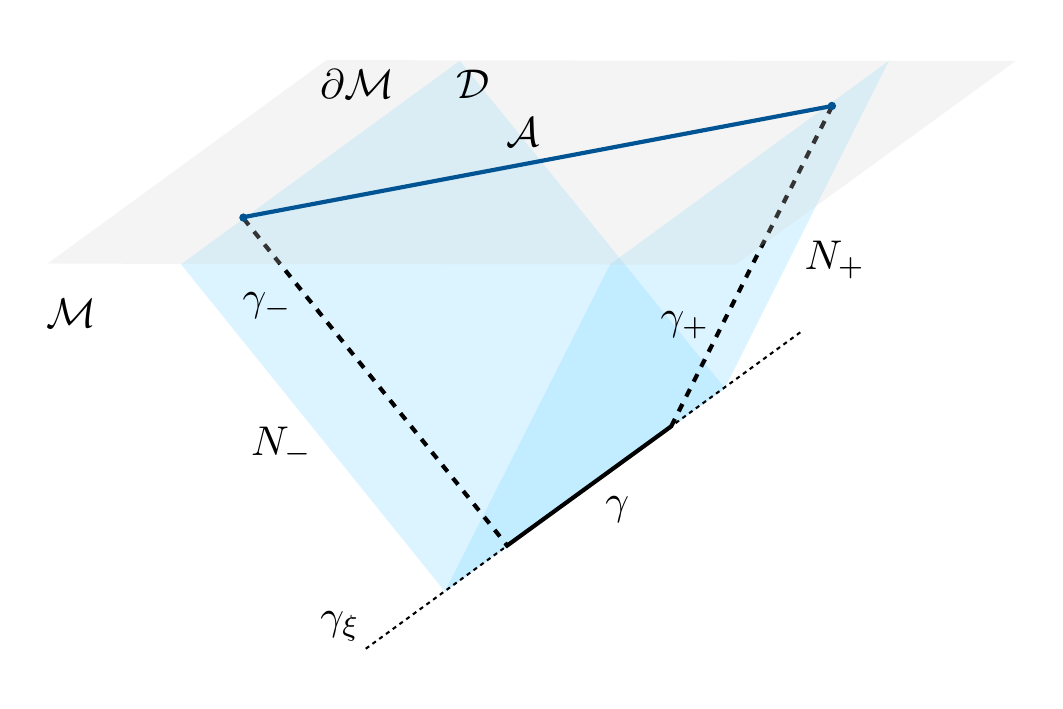} 
    \caption{In three dimensions the swing surface homologous to the boundary interval $\mA$ is given by $ \gamma_{\mA} = \gamma_- \cup  \gamma \cup \gamma_+$ where $\gamma_{\pm}$ are null geodesics lying on the light-sheets $N_{\pm}$ (blue) while $\gamma$ is the portion of the bifurcating surface $\gamma_\xi$ that connects $\gamma_+$ to $\gamma_-$ in the bulk. }
\label{s2:swingsurface} 
\end{figure}

Let us now describe how the bulk modular flow generator can be used to determine the RT/HRT surface in AdS/CFT and its generalization in models of non-AdS holography. In the AdS/CFT correspondence, the surface $\p\mA$ is the set of fixed points of both the bulk and boundary modular flow generators. This means that $\p\mA \subset \gamma_\xi$ and that the extremal surface $\gamma_\xi$ extends all the way from $\p\mA$ at the boundary into the bulk. In other words, $\gamma_\xi$ is just the RT/HRT surface. In contrast, in known examples of non-AdS holography including flat holography and the (W)AdS$_3$/CFT correspondence, the boundary endpoints $\p\mA$ are only fixed points of the boundary modular flow. As a result, $\p\mA\not\subset\gamma_\xi$ and the extremal surface $\gamma_\xi$ cannot be smoothly attached to a finite interval at the boundary. In this case, the bulk modular flow generator $\xi$ moves each point $p\in\p \mA$ along a null geodesic $\gamma_{(p)} \subset N_+\cup N_-$ that ends on $\gamma_\xi$. The surface that captures the entanglement entropy of subregion $\mA$ at the boundary is dubbed a {\it swing surface} and is defined by (see Fig.~\ref{s2:swingsurface})
\eq{
\gamma_\mA = \gamma \cup \gamma_{b\p}, \qquad \mathrm{where} \qquad   \gamma_{b\p}\equiv \cup_{p\in\p\mA}\gamma_{(p)}, \quad \p\gamma\equiv\gamma_\xi\cap\gamma_{b\p};\label{s2:swingsurfacedef}
}
where $\gamma_{b\p}$ are the {\it ropes} of the swing surface --- a collection of null geodesics tangent to the bulk modular flow that connect a point $p\in\p\mA$ at the boundary to a point $\gamma_\xi$ in the bulk; while $\gamma$ is the {\it bench} of the swing surface, which corresponds to the subset of fixed points of the bulk modular flow generator whose boundary $\p \gamma$ is the intersection of $\gamma_{b\p}$ with $\gamma_\xi$.

By construction, the swing surface \eqref{s2:swingsurfacedef} is homologous to the interval at the boundary and plays the role of the RT/HRT surface in non-AdS holography. 
In known models of non-AdS holography such as flat$_3$/BMSFT and (W)AdS$_3$/WCFT, the ropes $\gamma_{b\p}$ of the swing surface consist of two null geodesics $\gamma_+$ and $\gamma_-$ that connect the endpoints $\p\mA$ of the interval at the boundary to a spacelike geodesic in the bulk, namely to the bench of the swing surface. The entanglement entropy of the boundary interval $\mA$ is then given by the area of the swing surface, namely~\cite{Song:2016gtd,Jiang:2017ecm}
\be
S_\mA =\frac{{\rm Area} (\gamma_{\cal A} ) }{ 4G}, \label{proposal}
\ee
where $G$ is Newton's constant. In a companion paper~\cite{Apolo:2020bld}, we extend this geometric description of entanglement entropy beyond the vacuum and other highly symmetric states to cases where the modular flow is no longer realized geometrically.


\subsection{Entanglement and the equations of motion} \label{se:generalfirstlaw}

The bulk modular flow generator \eqref{s2:xi} allows us to express the modular Hamiltonian at the boundary as a gravitational charge in the bulk. Then, assuming the gravitational equations of motion hold, we provide a holographic derivation of the first law of entanglement entropy. Conversely, assuming the first law holds, we show how the equations of motion can be derived from entanglement.

The gravitational charge associated with the modular flow generator $\xi$ can be calculated using the covariant phase space formalism~\cite{ Wald:1993nt,Iyer:1994ys}, the latter of which has been successfully generalized to spacetimes that are not asymptotically AdS in~\cite{Barnich:2001jy, Guica:2008mu, Compere:2008cv} (see~\cite{Harlow:2019yfa} for recent developments). In the covariant phase space formalism, the infinitesimal variation of the Hamiltonian generating a diffeomorphism along an arbitrary vector field $\eta$ in a spatial region $\Sigma$ can be written as 
\eq{
 \delta { \mathscr H}_\eta^{\Sigma}[\phi] = \int_{\Sigma} {\bm \omega}[\delta_\eta \phi, \delta \phi] , \label{s2:ham}
}
where $\phi$ collectively denotes the fields of the theory and $\delta\phi$ denotes infinitesimal perturbations around $\phi$. The symplectic $d$-form $\bm \omega[\delta_1\phi, \delta_2\phi] =\delta_1 \bm \Theta[\phi, \delta_2\phi] -\delta_2 \bm \Theta[\phi, \delta_1\phi]$ is bilinear in $\delta_1\phi$ and $\delta_2\phi$ and defined in terms of the presymplectic potential $\bm\Theta[\phi, \delta \phi]$. The latter is determined from the Lagrangian describing the bulk theory of gravity via $\delta \bm L=\bm {E}\delta \phi+d\bm\Theta[\phi, \delta \phi]$ where $ {\bm E}$ denotes the bulk equations of motion. In a region with no topological obstructions, we can find a $(d-1)$-form $\bm\chi[\phi, \delta\phi]$ such that~\cite{Faulkner:2013ica}
\eq{
{\bm\omega}[\delta_\eta \phi, \delta\phi] = d \bm \chi_{\eta}[\phi, \delta\phi]+ 2 \delta E^g_{\mu\nu}  \eta^\mu\bm\epsilon^\nu,
}
 where $\bm\epsilon^\mu = \frac{1 }{ d !} \epsilon_{\mu \mu_2 \dots \mu_{d+1}} dx^{\mu_2}\wedge \dots \wedge dx^{\mu_{d+1}}$ and $\delta E_{\mu\nu}^g$ are the linearized equations of motion. The explicit form of $\bm \chi_{\eta}[\phi, \delta\phi]$ in Einstein gravity is given in \eqref{chidef}. Using Stokes' theorem, the bulk integral~\eqref{s2:ham} can be written as an integral over the boundary $\p \Sigma$ of $\Sigma$ such that
\eq{
 \delta { \mathscr H}_\eta^{\Sigma}[\phi] - 2\int_\Sigma  \delta E^g_{\mu\nu}  \eta^\mu\bm\epsilon^\nu =\int_{\Sigma} d\bm\chi_{\eta}[\delta \phi, \phi] =\int_{\p\Sigma} \bm\chi_{\eta}[\delta \phi, \phi]. \label{zero}
}

Let us now consider the bulk modular flow generator $\xi$ defined in \eqref{s2:xi}. Since $\xi$ is an exact Killing vector, the symplectic form vanishes and the Hamiltonian $\delta \mathscr H_{\xi}^\Sigma$ in \eqref{zero} is exactly zero. The infinitesimal gravitational charge along a surface $\cal C$ is defined by
  \eq{
  \delta \mathcal{Q}^{\cal C}_{\xi}[\phi] \equiv \int_{\cal C} \bm\chi_{\xi}[\delta \phi,\phi].\label{chargedef}
  }
 The holographic dictionary instructs us to identify the  gravitational charge evaluated at the asymptotic boundary with the modular Hamiltonian, namely 
 \eq{
 \delta \EV{\mathcal H_{mod}} =  \delta \mathcal{Q}^{\mA}_{\xi}[\phi] = \int_{\mA} \bm \chi_{\xi}[\delta \phi,\phi]. \label{s2:dHQb}
 }
This states that the expectation value of the vacuum modular Hamiltonian on an excited state dual to $\delta \phi$ is calculated by the infinitesimal gravitational charge associated with the modular flow generator $\xi$ on a background $\phi$ dual to the vacuum. The holographic dictionary \eqref{s2:dHQb} will be explicitly verified in flat$_3$/BMSFT and (W)AdS$_3$/WCFT in Sections~\ref{se:bmsft} and~\ref{se:wcft}, respectively. On the other hand, and in analogy with Wald's derivation of black hole entropy~\cite{Wald:1993nt}, we expect the holographic entanglement entropy of subregion $\mA$ to be identified with the gravitational charge evaluated on the swing surface $\gamma_{\mA}$~\eqref{s2:swingsurfacedef}
\eq{
\delta S_{\mA} = \delta \mathcal{Q}^{\gamma_\mA}_{\xi}[\phi]\equiv 
\int_{\gamma_{\mA}} \bm \chi_{\xi}[\delta \phi,\phi] . \label{s2:dsQA}
}
In~\cite{Apolo:2020bld} we propose a general prescription for holographic entanglement entropy based on swing surfaces where we show that  \eqref{s2:dsQA} is indeed satisfied in Einstein gravity.

So far we have kept the perturbations $\delta\phi$ off shell. In what follows, we describe how the bulk equations of motion imply the first law of entanglement entropy and vice versa.

\bigskip\noindent{\bf First law of entanglement from the bulk equations of motion.} If we assume that $\delta\phi$ satisfies the linearized equations of motion in the bulk, the left hand side of \eqref{zero} vanishes, implying that $d \bm\chi_\xi[\phi,\delta \phi]=0$. According to Stokes' theorem, the infinitesimal charge $\int_{\cal C} \bm\chi_{\xi}[\delta \phi,\phi]$ along a surface $\cal C$ is independent of $\cal C$ as long as the new integration surface is homologous to the old one. This implies, in particular, that the surface charge evaluated along the interval $\mA$ at the asymptotic boundary equals the one defined on the swing surface $\gamma_\mA$
\eq{
 \int_{\mA} \bm \chi_{\xi}[\delta \phi,\phi]=\int_{\gamma_{\mA}} \bm\chi_{\xi}[\delta \phi,\phi],
}
which, using the dictionary~\eqref{s2:dHQb} and~\eqref{s2:dsQA}, leads to the first law of entanglement entropy. Thus, assuming that the bulk equations of motion hold, we find that the first law of entanglement is satisfied as a consequence of Stokes' theorem, the holographic dictionary in eqs.~\eqref{s2:dHQb} and~\eqref{s2:dsQA}, and the fact that $\xi$ is an exact Killing vector of the background,
namely
\eq{
E^g_{\mu\nu}=0 \quad \Longrightarrow \quad \delta \EV{\mathcal H_{mod}}= \delta S_\mA.  \label{s2:firstlaw}
}

\bigskip\noindent{\bf Towards the linearized Einstein equations from entanglement.} We now reverse the logic and try to derive the linearized Einstein equations from entanglement. Let us assume that the holographic dictionary for the modular Hamiltonian \eqref{s2:dHQb} and for the entanglement entropy \eqref{s2:dsQA} are both valid. Then, the first law of entanglement entropy can be rewritten as the statement that the gravitational charges evaluated on $\mA$ and $\gamma_{\mA}$ are equal, namely
\eq{
\delta \EV{\mathcal H_{mod}} =  \delta S_{\mA} \quad \Longrightarrow \quad  \int_{\mA}  \bm{\chi}_\xi [g,\delta g] -\int_{\gamma_{\mA}}  \bm{\chi}_\xi[g,\delta g]= \int_{\p\Sigma}  \bm{\chi} _\xi[g,\delta g] =0, \label{ee1}
}
where $\p \Sigma = \mA \cup  \gamma_{\mA}$ is a closed surface.
Using Stokes' theorem and eq.~\eqref{zero} we then find
\eq{
\int _\Sigma  \delta E^g_{\mu\nu}  \xi^\mu\bm\epsilon^\nu=0, \label{deltaee}
}
where we used the fact that $\delta { \mathscr H}_\xi^{\Sigma}[\phi]=0$ when $\xi$ is an exact Killing vector. From this expression, one can try to obtain the local version of the bulk equations of motion in analogy with the analysis of \cite{Faulkner:2013ica} (see \cite{Godet:2019wje} for a similar  analysis  in flat holography).


\section{Flat$_3$/BMSFT} \label{se:bmsft}

In this section we study a three-dimensional toy model of holography for asymptotically flat spacetimes. The asymptotic symmetries of three-dimensional gravity in the absence of a cosmological constant are described by an infinite-dimensional symmetry algebra --- the three-dimensional version of the BMS algebra~\cite{Bondi:1962px,Sachs:1962wk,Sachs:1962zza,Barnich:2006av}. This motivates the conjecture that gravity in asymptotically flat three-dimensional spacetimes is dual to a BMS-invariant field theory (BMSFT) at the boundary~\cite{Barnich:2010eb,Bagchi:2010eg}. Supporting evidence for and related studies in this so-called flat$_3$/BMSFT correspondence include reproducing thermal entropy in the bulk from a Cardy-like formula at the boundary~\cite{Barnich:2012xq,Bagchi:2012xr}, entanglement entropy~\cite{Bagchi:2014iea,Basu:2015evh,Hosseini:2015uba,Jiang:2017ecm,Hijano:2017eii,Wen:2018mev,Godet:2019wje,Fareghbal:2019czx},  correlation functions~\cite{Bagchi:2015wna}, geodesic Witten diagrams~\cite{Hijano:2018nhq}, one-loop partition functions~\cite{Barnich:2015mui}, and energy conditions~\cite{Grumiller:2019xna} (see~\cite{Bagchi:2016bcd} for a review on BMS-invariant field theories and~\cite{Barnich:2012aw,Bagchi:2012cy,Fareghbal:2013ifa,Krishnan:2013wta,Fareghbal:2014oba,Barnich:2014zoa,Fareghbal:2014qga,Fareghbal:2015bxd,Bagchi:2016geg,Bagchi:2019unf,Merbis:2019wgk} for related studies).  

In this paper, we will explore further aspects of entanglement entropy in flat holography and study in detail the modular Hamiltonian in both the bulk and boundary sides of the correspondence, discuss the bound on modular chaos, and derive the first law of entanglement entropy. 


\subsection{A brief introduction to flat$_3$/BMSFT}

In this section we review holography for three-dimensional flat spacetimes focusing on aspects related to symmetry.

\subsubsection{Three-dimensional gravity in flat space}

Let us consider three-dimensional Einstein gravity in asymptotically flat spacetimes satisfying the following boundary conditions near future null infinity~\cite{Barnich:2010eb}
  \eq{
  \begin{gathered}
    g_{rr} = 0, \qquad g_{ru} = - 1 + {\cal O}(1/r), \qquad g_{r\phi} = 0, \qquad g_{u\phi} = {\cal O}(1),   \\
  g_{uu} = {\cal O}(1), \qquad g_{\phi\phi} = r^2,
  \end{gathered}\label{flatbc}
  }
where $(u, r, \phi)$ denote retarded Bondi coordinates and $\phi \sim \phi + 2\pi$. In Bondi gauge, the space of solutions to Einstein's equations compatible with the boundary conditions~\eqref{flatbc} can be parametrized by two periodic functions $\Theta(\phi)$ and $\Xi(\phi)$ such that
\eq{
ds^2 = g_{\mu\nu} dx^{\mu} dx^{\nu} = h_{uu} du^2-2 du dr+2 h_{u\phi}du d\phi+r^2 d\phi^2, \label{Flatmetric}
}
where $h_{uu}$ and $h_{u\phi}$ are given by
\be\label{huuphi}
h_{uu}=\Theta(\phi), \qquad h_{u\phi}=\Xi(\phi)+\frac12 u\partial_\phi\Theta(\phi).
\ee

The boundary conditions~\eqref{flatbc} lead to an infinite-dimensional asymptotic symmetry algebra that is generated by the asymptotic Killing vector
\be
\eta = \big[ \sigma(\phi) + u \epsilon'(\phi)\big] \partial_u + \epsilon(\phi) \partial_\phi - r \epsilon'(\phi)  \p_r + \mathrm{subleading}, \label{flatkillings}
\ee
where $\epsilon(\phi)$ and $\sigma(\phi)$ are two arbitrary functions and $\sigma'(\phi) \equiv \partial_{\phi} \sigma(\phi)$. The conserved charges associated with~\eqref{flatkillings} are integrable in the space of solutions and are given by
  \eq{
  {\cal Q}_{\epsilon_n} &= \frac{1}{8\pi G} \int d\phi \,e^{i n\phi} \Xi(\phi), \qquad {\cal Q}_{\sigma_n} = \frac{1}{16\pi G} \int d\phi\, e^{i n \phi} \Theta(\phi), \label{flatcharges}
  }
where we expanded $\epsilon(\phi)$ and $\sigma(\phi)$ in Fourier modes, i.e.~$\epsilon(\phi) = \sum_n \epsilon_n e^{i n\phi}$ and $\sigma(\phi) = \sum_n \sigma_n e^{i n \phi}$.  It is not difficult to verify that the gravitational charges~\eqref{flatcharges} satisfy the three-dimensional BMS algebra with central extensions $c_L = 0$ and $c_M = 3/G$~\cite{Barnich:2006av} 
  \eqsp{
  [{\cal Q}_{\epsilon_m}, {\cal Q}_{\epsilon_n}] &= (m - n) {{\cal Q}_{\epsilon_{m+n}}} + \frac{c_L}{12} m(m^2 -1) \delta_{m+n,0}, \\
  [{\cal Q}_{\epsilon_m}, {\cal Q}_{\sigma_n}] &= (m - n) {\cal Q}_{\sigma_{m+n}} + \frac{c_M}{12} m(m^2 -1) \delta_{m+n,0},\\
  [{\cal Q}_{\sigma_m}, {\cal Q}_{\sigma_n}] &= 0. \label{3dbms}
  }
This algebra corresponds to an In\"on\"u-Wigner contraction of the Virasoro algebra that can be obtained by taking the $\ell \to \infty$ limit of gravity on AdS$_3$ backgrounds satisfying Brown-Henneaux boundary conditions where $\ell$ is the scale of AdS.

We are especially interested in the zero-mode solutions described by \eqref{Flatmetric} with $\Theta(\phi) = M$ and $\Xi(\phi) = J/2$.  Up to factors of $8G$, the parameters $M$ and $J$ correspond to the canonical energy and the angular momentum of the background spacetime. In particular, $M  = -1$ and $J = 0$ corresponds to the global Minkowski vacuum while the zero point energy is usually taken at the null orbifold where $M = J = 0$. Solutions with $M>0$ have a Cauchy horizon and are usually referred to as flat cosmological solutions (FCS). We are also interested in the vacuum in flat Poincar\'e coordinates which is described by
\eq{
ds^2= -2 dudr +r^2 d\zc^2, \label{flatPoincare}
}
where $\zc \in(-\infty, \infty)$. This is the flat limit of the Poincar\'e patch of AdS obtained by sending the AdS radius to infinity such that its boundary is a plane. Certain computations simplify in this gauge, but these can be straightforwardly generalized to general zero-mode backgrounds on the cylinder (see Sec~\ref{se:zeroMode} for details).

Locally, the zero-mode backgrounds with $\Theta(\phi) = M$ and $\Xi(\phi) = J/2$ feature six independent Killing vectors that can be written as
\eqsp{
L_j&=  \Big[ \Big(u+\frac{J }{ 2 M}\phi \Big)\epsilon_j' -\frac{J}{2 M} \epsilon_j   \Big] \partial_u +\Big(\epsilon_j - \frac{1}{ r} \p_\phi L_j^u\Big) \p_\phi  - \Big(\frac{J}{2r} \p_\phi L_j^u + r \partial_\phi L_j^\phi \Big) \p_r, \\
M_j&=  \sigma_j \p_u  -\frac{1}{r}\sigma_j'\p_\phi+\Big(\sigma_j''-\frac{J}{2r} \sigma'_j \Big)\p_r ,\label{BulkKilling}
}
where $j \in\{0,\pm 1\}$ while $\epsilon_j$ and $\sigma_j$ are functions of $\phi$ (or $z$ for the Poincar\'e vacuum) that depend on the background spacetime. For zero-mode backgrounds with $M \ne 0$, $\epsilon_j$ and $\sigma_j$ read
\eq{\label{BulkKillingMJ}
M \neq 0 &:   \quad \epsilon_j =\frac{1}{\sqrt{M}}e^{-j  \sqrt{M}\phi},   \qquad \sigma_j  =-\frac{1}{\sqrt{M}}e^{-j  \sqrt{M}\phi},
}
while for the Poincar\'e vacuum \eqref{flatPoincare} they are instead given by
\eq{
M=J=0&:   \quad   \epsilon_j =  -\zc^{j+1},  \qquad \sigma_j =-\zc^{j+1}.\label{BulkKillingplane}
}
The Killing vectors \eqref{BulkKilling} satisfy the Poincar\'{e} algebra under the Lie bracket, which is the global part of the BMS algebra given in \eqref{3dbms}. 


\subsubsection{BMS-invariant field theories}

Three-dimensional gravity on asymptotically flat spacetimes is conjectured to be dual to a BMS-invariant field theory (BSMFT) at the boundary~\cite{Barnich:2010eb,Bagchi:2010eg}. The latter corresponds to a class of ultrarelativistic quantum field theories that are invariant under spacetime symmetries obtained by reparametrizations of the form
\eq{
\tilde \phi=f(\phi), \qquad  \tilde u=u f'(\phi)+g(\phi), \label{flatcoordT}
}
where $f(\phi)$ and $g(\phi)$ are arbitrary functions and $f'(\phi) \equiv \p_{\phi} f(\phi)$. In particular, the $u$ and $\phi$ components of the generator of the coordinate transformation~\eqref{flatcoordT} agree with the corresponding components of the asymptotic Killing vector \eqref{flatkillings} on the bulk side of the flat$_3$/BMSFT correspondence.

BMS-invariant field theories are equipped with a pair of conserved currents $\mathcal J(\phi)$ and $\mathcal P(\phi)$ that generate the local coordinate transformations~\eqref{flatcoordT}. Under the coordinate transformation~\eqref{flatcoordT} these currents transform as
\eqsp{ 
\tilde{ \mathcal P}(\tilde{\phi}) &= (\tilde f')^{2}\,  \mathcal P(\phi) +  \frac{c_{{M}}}{12}\{\phi, \tilde \phi \},   \\ \tilde{\mathcal J}({\tilde \phi}) &=  (\tilde f')^{2} \,   \mathcal J(\phi)  + 2  \tilde  f'\,   \tilde  g'  \, \mathcal P(\phi) +   (\tilde f')^{2}\,  \tilde g\, \mathcal P'(\phi) + \frac{c_{{L}}}{12}  \{\phi, \tilde \phi \} + \frac{c_{{M}}}{12} [\![  ( \tilde f , \tilde g  ),  \tilde\phi ]\!],\label{chargeTsf}
}
where $\phi = \tilde f(\tilde \phi)$ and $u = \tilde u \tilde f'(\tilde \phi) + \tilde g(\tilde\phi)$ denote the inverse of the coordinate transformations \eqref{flatcoordT} and we have dropped the $\tilde{\phi}$ dependence of $\tilde f(\tilde \phi)$ and $\tilde g(\tilde \phi)$ for convenience. In \eqref{chargeTsf}, $\{\cdot ,\cdot \}$ denotes the ordinary Schwarzian derivative, namely $\{\tilde f,\tilde\phi \} = {\tilde f''' }/{\tilde f' } - ({3}/{2}) ({\tilde f'' }/{\tilde f' })^2$, while $[\![ \cdot , \cdot ]\!]$ denotes the ``BMS  Schwarzian'' which is given by
\be
[\![  (  \tilde f , \tilde g ),  \tilde\phi ]\!]  =\frac{ \big[3 (\tilde f'')^{2}-\tilde f' \tilde f'''    \big]\tilde g'  -3 \tilde f' \tilde f'' \tilde g''+(\tilde f')^2 \tilde g'''
}{(\tilde f' )^3  } .
\ee
One way to derive \eqref{chargeTsf} is to consider the transformation rules of the stress tensor in two-dimensional CFTs and then perform an \.In\"on\"u-Wigner contraction to arrive at the transformation rules for the currents of BMSFTs.\footnote{This step was done in~\cite{Basu:2015evh}. To arrive at the transformation rules of ${\cal P}(\phi)$ and ${\cal J}(\phi)$ we need to take eq.~(21) there and replace:  $t\rightarrow  \phi$, $x\rightarrow u $, $T^{(2)} \rightarrow {\mathcal P}(\phi)$, and $T^{(1)}  \rightarrow  {\mathcal J}(\phi)+u{\mathcal P}'(\phi)$.} In particular, the BMS Schwarzian arises from the expansion $ \{\tilde f + \epsilon\, \tilde g  ,\tilde\phi \} = \{\tilde f,\tilde\phi \} + \epsilon \; [\![  (\tilde f, \tilde g),   \tilde\phi ]\!]  +\mathcal O(\epsilon^2)$. Alternatively, eq.~\eqref{chargeTsf} can be deduced from the infinitesimal transformation laws of the currents obtained directly from the BMS algebra~\eqref{3dbms}. 

Finally, the Fourier modes of the $\mathcal J(\phi)$ and $\mathcal P(\phi)$ currents are respectively given by 
\eq{  
\mathcal L_n = - \frac{1}{2\pi} \int d\phi e^{in\phi} \mathcal J(\phi), \qquad \mathcal M_n = -\frac{1}{2\pi} \int d\phi e^{in\phi} \mathcal P(\phi), \label{cylindercharges}
} 
and satisfy the same BMS algebra of asymptotically flat spacetimes \eqref{3dbms} where \eq{
{\cal Q}_{\epsilon_n} \to {\cal L}_n, \qquad {\cal Q}_{\sigma_n} \to {\cal M}_n.
} 
The matching of the bulk and boundary symmetry algebras provides a first indication of holography for asymptotically flat three-dimensional spacetimes. In particular, from eq.~\eqref{cylindercharges}, we deduce the holographic dictionary relating the bulk $\{\Xi(\phi), \Theta(\phi)\}$ and boundary $\{\mathcal J(\phi),\mathcal P(\phi)\}$ variables, namely 
\eq{ 
\Xi(\phi)= - 4 G  \mathcal J(\phi), \qquad \Theta(\phi) = -8 G  \mathcal P(\phi). \label{dictionary} 
} 
Finally, we note that \eqref{dictionary} is compatible with the holographic dictionary between the bulk metric and the boundary stress tensor obtained in \cite{Fareghbal:2013ifa}.
 

\subsection{The boundary modular Hamiltonian} \label{flatse:bdymodham}

In this section we derive the vacuum modular Hamiltonian for a single interval in BMSFTs defined on the plane. In particular, we find a general expression for the modular flow generator by exploiting the global symmetries of BMSFTs and show that the bound on modular chaos is satisfied in this class of ultrarelativistic field theories.

\subsubsection{Modular flow for a single interval} \label{flatse:modflow}
 
We begin by considering a general interval $\mA$ on the vacuum state with endpoints 
\eq{
\p \mA=\{ (u_-,\zc_-),(u_+,\zc_+)\}, \qquad l_u \equiv u_+ - u_-, \quad l_\zc \equiv \zc_+ - \zc_-, \label{interval0}
}
where $(u,\zc)$ denote the coordinates on the plane. The modular flow generator $\zeta$ for the vacuum (and other highly symmetric states) can be derived from the generalized Rindler transformation or the general prescription described in Section~\ref{se:modularH}. Let us consider the second approach and try to find a vector that satisfies the conditions listed towards the end of Section~\ref{se:modularH}. The first condition tells us that we can write the modular flow generator as a linear combination of vacuum symmetry generators
\be
\zeta =\zeta^u \p_u + \zeta^\zc \p_\zc = \sum_{j=-1}^1 (a_j   \ell_j +b_j    m_j ), \label{flatmodflowgen}
\ee
where $a_j$ and $b_j$ are coefficients to be determined, while $\ell_j$ and $m_j$ with $j \in \{-1,0,1\}$ denote the vacuum symmetry generators on the plane
\eq{ 
\ell _j & = -\zc^{j+1}\partial_\zc-(j+1)\zc^j u \partial_u, \qquad m_j = { -} \zc^{j+1}\partial_u, \label{BMSgenerator}
}
 which satisfy the BMS algebra
 \eq{ 
[ \ell_j, \ell_k] =  (j-k) \ell_{j+k}, \qquad  [ \ell_j, m_k]  = (j-k)  m_{j+k}, \qquad [  m_j,  m_k]  = 0.  \label{BMSalgebra}
}

\begin{figure}[h] 
   \centering 
    \includegraphics[scale=0.33333]{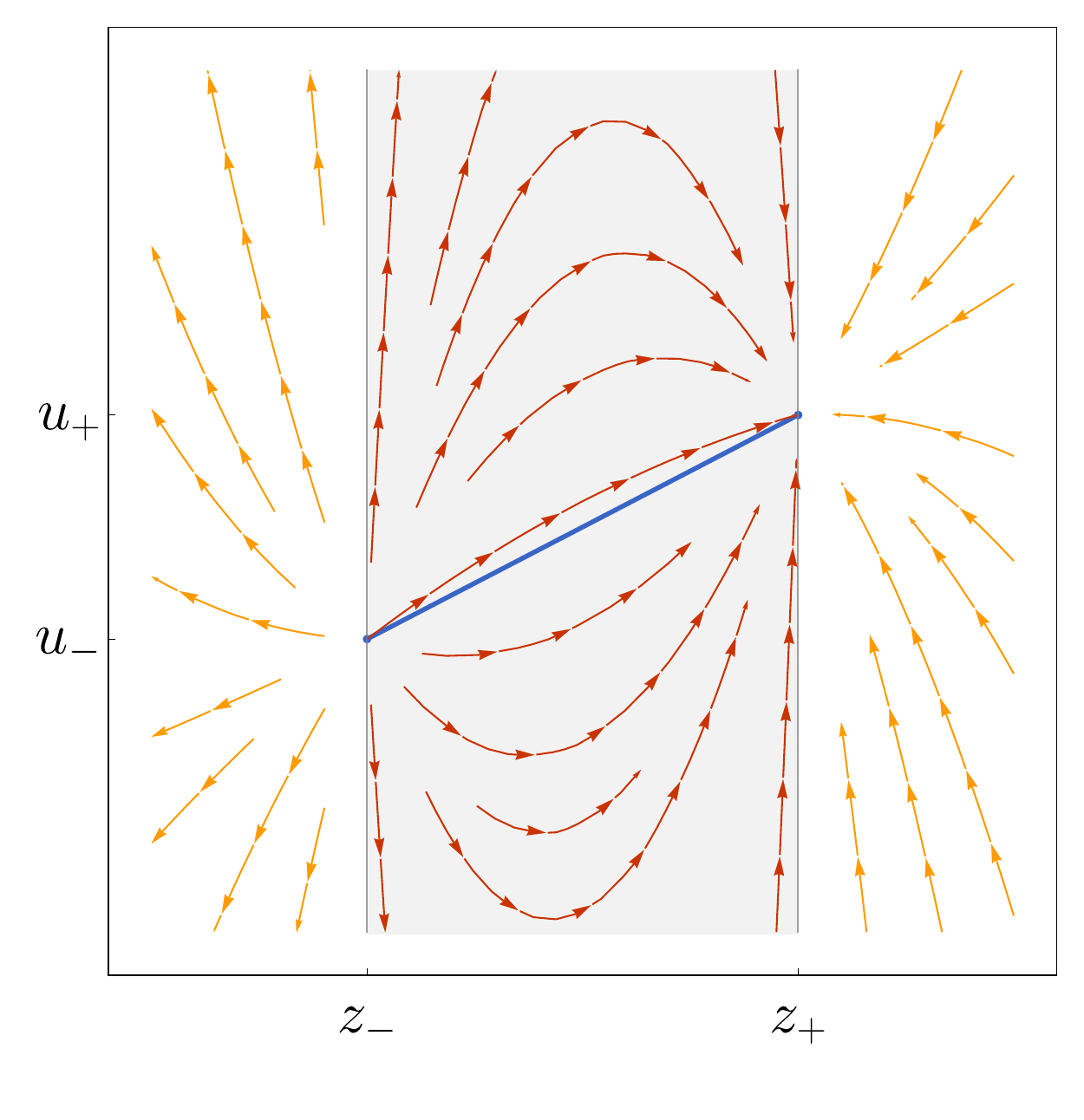}
    \caption{Modular flow (red) for a single interval (blue) in BMSFTs. The causal domain of dependence (gray) and the behavior of the modular flow near the endpoints differs from that of CFTs. We also show the modular flow generated by $\bar{\zeta}$ in orange.} \label{fig:flatmodflow}
    \end{figure}

The causal domain $\mathcal D$ of $\mA$ is illustrated in Fig.~\ref{fig:flatmodflow} and  consists of an infinite strip along the $u$ coordinate. The coefficients $a_j$ and $b_j$ in \eqref{flatmodflowgen} can be determined by requiring that the boundary $\p\cal D$ of the causal domain is invariant under the modular flow and that $e^{i\zeta}$ maps any point in $\cal D$ back to itself. More specifically, by requiring that the modular flow generator vanishes at the endpoints of the interval $\mA$ we find
\eqsp{ \label{abvalue}
(a_+,a_0,a_-) & = a_+ \big(1, -\zc_-   -\zc_+, \zc_+ \zc_- \big), \\
(b_+,b_0,b_-) & = b_+ \big(1, -\zc_-   -\zc_+, \zc_+ \zc_- \big) +a_+\big(0,-u_--u_+,u_+\zc_-+u_-\zc_+\big).
}
In order to determine the remaining $a_+$ and $b_+$ coefficients, we note that under the modular flow $e^{s\zeta}$ the trajectory $( \zc(s), u(s))$ obeys the following differential equations
\eq{
\p_s  u(s) =\zeta^u, \qquad \p_s \zc(s)= \zeta^\zc,
}
where $\zeta^u$ and $\zeta^\zc$ can be obtained from eq.~\eqref{flatmodflowgen}. The solution for $\zc(s)$ is given by
\be
\zc(s)=\frac{ e^{a_+ \zc_+ s +c_0 \zc_- } \zc_+ - e^{a_+ \zc_- s+ c_0 \zc_+ } \zc_-} {e^{a_+ \zc_+ s +c_0 \zc_- } - e^{a_+ \zc_- s+ c_0 \zc_+ }},
\ee
where $c_0$ is a constant, while the solution for $u(s)$ is not particularly illuminating and is therefore omitted. Requiring $\zc(s)= \zc(s+i)$ and $u(s)= u(s+i)$, we find that $a_+$ and $b_+$ are given, up to an overall sign, by
\eq{ 
a_+=\frac{2\pi}{\zc_+-\zc_-}, \qquad  b_+ = -  \frac{2\pi (u_+ - u_-)}{(\zc_+-\zc_-)^2 }.
}
In this way, the coefficients in~\eqref{flatmodflowgen} are completely determined and found to satisfy
\eqsp{
(a_+,a_0,a_-) & = \frac{2\pi}{\zc_+ - \zc_-}  \big(1  ,  -\zc_-   -\zc_+ , \zc_+ \zc_- \big), \\
(b_+,b_0,b_-) & = \frac{2\pi}{( \zc_+ - \zc_-)^2}  \big(u_- -u_+ ,   2u_+ \zc_- -2 u_- \zc_+ ,  u_- \zc_+^2-u_+ \zc_-^2  \big). \label{abs}
} 

The modular flow generator $\zeta$ for an arbitrary interval \eqref{interval0} can be written as
\eq{
 \zeta & =\sum_{j=-1}^1 (a_j \ell_j+b_j m_j)=\big[ T(\zc) +u Y'(\zc)  \big] \p_u +Y(\zc) \p_\zc, \label{modularflow}
}
where the functions $T(\zc)$ and $Y(\zc)$ are given by
\arraycolsep=5pt \renewcommand{\arraystretch}{.4}
\eq{
T(\zc)&= \left\{\begin{array}{lcl}
    \frac{2 \pi  [u_+ (\zc -\zc_-)^2-u_- (\zc -\zc_+)^2 ]}{(\zc_+-\zc_-)^2}, & & \zc \in [\zc_-, \zc_+], \\
    \\
    0 && \textrm{otherwise}, 
    \end{array} \right. \label{YT} \\
Y(\zc) &= \left\{ \begin{array}{lcl}
  -\frac{2 \pi  (\zc -\zc_-) (\zc -\zc_+)}{\zc_+-\zc_-},  \phantom{z-2\,z}&& \zc \in [\zc_-, \zc_+], \\
  \\
  0 && \textrm{otherwise}.
  \end{array} \right.\label{YT2}
}
The modular flow generator $\zeta$ specified by the functions \eqref{YT} and \eqref{YT2} is nonvanishing only in the causal domain of dependence $\cal D$ of $\mA$.
We can also define the modular flow generator $\bar\zeta$ for the complement of $\mA$. It is not difficult to verify that $\bar \zeta$ can be written in the form of \eqref{modularflow}, but with the functions $T(z)$ and $Y(z)$ in eqs.~\eqref{YT} and \eqref{YT2} nonvanishing only for $\zc\not\in  [\zc_-, \zc_+]$. 

We now show that \eqref{modularflow} is the same expression for the modular flow generator obtained from the generalized Rindler method in~\cite{Jiang:2017ecm}. The generalized Rindler transformation $(u, \zc) \to (\tilde u, \tilde \zc)$ maps the causal domain of dependence $\cal D$ of the interval $\cal{A}$ to a noncompact space and introduces a thermal identification for the tilded variables $( \tilde u,\tilde\zc) \sim  (\tilde u -\tilde{\beta}_u,\tilde\zc+\tilde{\beta}_\zc)$. The Rindler transformation is explicitly given by
\eqsp{
 \tanh \frac{\pi\tilde{\zc}}{ \tilde{\beta}_\zc} &= \frac{2 \zc-\zc_+-\zc_-}{ {l_\zc }},  \\
 {\tilde{u}}+\frac{{\tilde\beta}_u }{ {\tilde \beta}_\zc} \tilde{\zc} &= \frac{\tilde{\beta}_\zc\big[(2u- u_+  - u_- )l_\zc - (2\zc- \zc_+ - \zc_- )l_u \big]}{ \pi \big[ l_\zc^2 - (2\zc- \zc_+ - \zc_-)^2 \big]}, \label{rindlertNO}
}
where we have used $l_u \equiv u_+ - u_-$ and $l_\zc \equiv \zc_+ - \zc_-$. The modular flow generator $\zeta$ corresponds to the generator of this thermal identification
\eq{
\zeta&= {\tilde{\beta}_\zc} \p_{\tilde \zc} - {\tilde\beta}_u \p_{\tilde{u}},
 }
and agrees with \eqref{modularflow} after using the Rindler transformation \eqref{rindlertNO}.


\subsubsection{Modular chaos} \label{flatse:modularalgebra}
 
We now show that the modular flow generator satisfies the bound on modular chaos described in Section~\ref{se:modularalgebra}. A general shape deformation of the modular flow generator \eqref{modularflow} can be obtained by varying the endpoints $(u_\pm,\zc_\pm)$. It is not difficult to  show that varying the endpoints of the interval $\mA$ leads to a deformation of the modular flow generator that satisfies the eigenvalue problem~\eqref{s2:modularchaos2}, namely
\eq{
[\zeta, \delta_\pm\zeta]  =  \pm 2\pi \delta_\pm\zeta,  \qquad \delta_\pm\zeta=\delta u_\pm \p_{u_\pm}\zeta+\delta \zc_\pm \p_{\zc_\pm}\zeta\,.\label{scramblingBMSFT}
}
Geometrically, the eigenvector $\delta_+ \zeta$ moves the endpoint $(u_+,\zc_+)$ an amount $(\delta u_+, \delta \zc_+)$ while keeping the other endpoint $(u_-,\zc_-)$ fixed (similar statements hold for $\delta_- \zeta$). As a result, shape deformations of the interval $\mA$ with one endpoint fixed saturate the bound on modular chaos in BMSFTs and correspond to modular scrambling modes \cite{deBoer:2019uem}. This behavior differs from that of CFTs where the eigenvectors of the modular flow generator move the two endpoints of $\mA$ simultaneously along specific null directions. This difference is related to the behavior of the modular flow generator near the endpoints of the interval ${\cal A}$: in CFTs, the modular flow generator reduces to a boost near the endpoints; while in BMSFTs, it corresponds to the sum of two dilatations. We will  comment on the physical implication of this observation on the bulk side of the flat$_3$/BMSFT correspondence in Section~\ref{se:bulkflowandHEE}.


\subsubsection{The modular Hamiltonian}\label{flatse:modH}

We now have the necessary ingredients to derive the vacuum modular Hamiltonian for single intervals in BMSFTs defined on the plane.  Up to a constant, the modular Hamiltonian can be obtained by replacing the coordinate space generators \eqref{modularflow} with the corresponding conserved charges, namely $\ell_j \to {\cal L}_j$ and $m_j \to {\cal M}_j$. On the real plane with real $u$ and $z$ coordinates, the charges ${\cal L}_j$ and ${\cal M}_j$ are defined by
\eq{
\mathcal{L}_j &= \frac{1}{2\pi } \int d\zc  \, \zc^{j+1} \mathcal J(\zc), \qquad \mathcal{M}_j = \frac{1}{2\pi } \int d\zc \,  \zc^{j+1} \mathcal P(\zc).
}
Then the modular Hamiltonian is, up to a constant,  given by the charge associated with the modular flow generator \eqref{modularflow}
\eq{
\mathcal H_\zeta &= \sum_{j=-1}^1 (a_j {\cal L}_j+b_j {\cal M}_j)  =  -\frac{1}{2\pi} \int_{\zc_-}^{\zc_+} d\zc  \big[ T(\zc)\mathcal P(\zc)+ Y(\zc) \mathcal J(\zc) \big], \label{ModHam}
}
where we have used the fact that the $T(\zc)$ and $Y(\zc)$ functions \eqref{YT} are non-vanishing only within the causal domain $\cal D$.
 
As a consistency check, let us consider an alternative derivation of \eqref{ModHam}. Note that the ordinary Hamiltonian in Rindler space  can be written as %
\eq{ 
 {\tilde {\mathcal  H}}_\zeta  &   =  -\frac{1}{2\pi} \int d\tilde \zc  \Big[ \tilde \beta_\zc \tilde{\mathcal J}(\tilde \zc)-   \tilde \beta_u   \tilde{\mathcal P}(\tilde \zc) \Big]. \label{RindlerHam}
}
Using the finite BMS transformations of the currents $\mathcal J(\zc)$ and $\mathcal P(\zc)$ \eqref{chargeTsf} under the Rindler transformation \eqref{rindlertNO}, together with the derivative relations 
\eq{
 \tilde f'= \frac{1}{f'}, \qquad \tilde g =  -\frac{g}{f'}, \qquad \tilde g'=  \frac{gf''-g'f'}{f'{}^3},\label{fgInverse}
}
we find that the Rindler Hamiltonian \eqref{RindlerHam} indeed reproduces the expression given in eq.~\eqref{ModHam}, up to constant terms proportional to the central charges. These constant terms do not contribute to variations of the modular Hamiltonian that result from variations of the state and do not play a role in what follows. Another consistency check on the modular Hamiltonian is that it satisfies the first law of entanglement entropy as we will discuss in more detail later.
 

\subsection{Bulk modular flow and the first law} \label{flatse:bulkmodflow}

In this  section we extend the vacuum modular flow generator into the bulk and use it to determine the holographic entanglement entropy of an interval $\mA$ at the boundary. We also calculate the gravitational charge associated with the bulk modular flow generator using the covariant formulation of gravitational charges~\cite{Wald:1993nt,Iyer:1994ys, Barnich:2001jy}. In particular, we show that this charge agrees with the modular Hamiltonian of BMSFTs and, following the discussion in Section~\ref{se:generalfirstlaw}, provide a holographic derivation of the first law of entanglement entropy.

In this section we work with the Poincar\'e vacuum \eqref{flatPoincare} which is the spacetime dual to the vacuum of BMSFTs defined on the plane. The discussion is still valid for more general zero-mode backgrounds as described in  Section~\ref{se:zeroMode}.

\subsubsection{Bulk modular flow and holographic entanglement} \label{se:bulkflowandHEE}

Let us begin by extending the modular flow generator into the bulk which is used to give a bulk interpretation of the modular scrambling modes discussed in Section~\ref{flatse:modularalgebra} and rederive the holographic entanglement entropy of single intervals in flat$_3$/BMSFT~\cite{Jiang:2017ecm}.

The boundary modular flow generator~\eqref{modularflow} can be extended into an exact Killing vector $\xi$ in the bulk by requiring that $\xi\big|_{\p {\cal M}} = \zeta$. This can be accomplished by substituting the boundary symmetry generators \eqref{BMSgenerator} with the bulk generators of the Poincar\'e vacuum given in eqs.~\eqref{BulkKilling} and \eqref{BulkKillingplane}, namely $\ell_i\to L_i$ and $m_i\to M_i$, such that 
\eq{
 \xi &=\sum_{j=-1}^1 (a_j L_j+b_j M_j), \label{bulkmodflow}
   }
 where the coefficients $a_j$ and $b_j$ are given in \eqref{abs}. More explicitly, the components of the bulk modular flow generator can be written as 
 \eq{
 \xi^u &=  T(\zc) + u  Y'(\zc), \qquad  \xi^\zc = Y(\zc) -\frac{u}{r} Y''(\zc)-\frac{1}{r} T'(\zc), \qquad \xi^r =  -r \partial_\zc \xi^\zc, \label{bulkmodflowcomponents}
}
where the functions $T(\zc)$ and $Y(\zc)$ are given in \eqref{YT}. 

We note that the modular flow generator $\xi$ has a bifurcating Killing horizon
\eq{
N_\pm: \, r=- \frac{u-u_\pm }{ (\zc-\zc_\pm)^2}, \label{horizon}
}
and the bifurcating surface $\gamma_\xi$, which is the intersection of $N_+$ and $N_-$, is given  by
\eq{
\gamma_\xi: \, u=\frac{l_u}{4l_{\zc}} (2\zc - \zc_+ - \zc_-) - \frac{l_{\zc}^2 }{ 8}r, \qquad r = -\frac{2 l_u}{( 2\zc - \zc_+ - \zc_- ) l_\zc}. \label{RTsurf}
}
It is not difficult to show that $\gamma_\xi $ is the set of fixed points of the bulk modular flow generator $\xi$ and that the latter satisfies eqs.~\eqref{normaleq1} and~\eqref{normaleq2}, namely
\be\label{Npm}
\xi_{[\mu} \nabla_\nu \xi_{\lambda]}\big|_{N_\pm}=0, \qquad \xi^\nu \nabla_\nu \xi^\mu \big|_{N_\pm} = \pm 2\pi \xi^\mu.
\ee

\bigskip
\noindent{\bf Modular scrambling modes.} Let us comment on the modular scrambling modes \eqref{scramblingBMSFT} discussed in the context of the bound on modular chaos in Section~\ref{flatse:modularalgebra}. We can extend $\delta_\pm \zeta$ into the bulk by replacing the boundary symmetry generators by the corresponding generators in the bulk such that
\eq{
\delta_\pm \xi&=\sum_{j=-1}^1 (\delta u_\pm \p_{u_\pm}+\delta \zc_\pm \p_{\zc_\pm})(a_j L_j+b_j M_j) =\delta u_\pm \p_{u_\pm}\xi+\delta \zc_\pm \p_{\zc_\pm}\xi.
}
It follows that these variations of the bulk modular flow generator satisfy a similar equation to  \eqref{scramblingBMSFT}, namely
\eq{
[\xi, \delta_\pm \xi ]  =  \pm 2\pi  \delta_\pm \xi,
}
where $[\cdot, \cdot]$ is the Lie bracket. Furthermore, we find that $\delta_\pm \xi$ satisfy
\eq{
 \delta_+ \xi \cdot \delta_+\xi \big|_{N_-} = \delta_- \xi \cdot \delta_-\xi  \big|_{N_+}=0, \qquad \delta_+ \xi \cdot \delta_-\xi \big|_{\gamma}\neq0.
 }
As a result, $\delta_\pm \xi $  generates a deformation of $\gamma_\xi$ along $N_{\mp}$, or equivalently, a deformation of $N_\pm$. From \eqref{horizon} we learn that $N_+$ depends only on the  $(u_+,\zc_+)$ endpoint. Thus, a deformation of $N_+$ only changes the position of $(u_+,\zc_+)$ while it leaves the other endpoint $(u_-,\zc_-)$ unchanged. This is consistent with the behavior of $\delta_+ \zeta$ on the BMSFT side of the correspondence as described in Section~\ref{flatse:modularalgebra}. Note that the behavior of the modular scrambling modes in flat$_3$/BMSFT is different from that in the AdS/CFT correspondence. There, a null deformation of the RT/HRT surface along $N_+$ or $N_-$ moves the two endpoints simultaneously along opposite null directions, a picture that is consistent with the behavior of the modular scrambling modes in CFT.

\bigskip
\noindent{\bf Holographic entanglement entropy.} As described in Section~\ref{se:genrt}, the bulk modular flow generator~\eqref{bulkmodflow} allows us to determine the location of the swing surface whose area yields the entanglement entropy of the field theory at the boundary. We first note that the modular flow generator is null and proportional to $\p_r$ at the endpoints $(u,\zc) =(u_\pm,\,\zc_\pm)$ of the interval $\mA$. The ropes $\gamma_{\pm}$ of the swing surface correspond to null geodesics emanating from the endpoints of $\mA$ which are parallel to the modular flow generator such that
\eq{
 \gamma_\pm: \, u = u_\pm, \quad \zc = \zc_\pm.   \label{gammaplusminus}
}
These null lines intersect the bifurcating surface $\gamma_{\xi}$ \eqref{RTsurf} at the points $\p\gamma_\pm = \gamma_\xi \cap \gamma_\pm$ which are given by
\eq{
\p\gamma_\pm: \Big(u= u_\pm , \zc= \zc_\pm, r= \mp\frac{2l_u}{l_\zc^2} \Big). \label{endpoints}
}
 Let $\gamma$ denote the segment of $\gamma_\xi$ lying between the ropes $\gamma_\pm$ which is bounded by the points $\p\gamma_\pm$ (see Fig.~\ref{s2:swingsurface}). The swing surface $\gamma_\mA$ is then given by the union of the null ropes $\gamma_\pm$ and the spacelike bench $\gamma$, namely
  \eq{
  \gamma_{\cal A} = \gamma_- \cup  \gamma \cup \gamma_+. \label{swgsurf}
  }
The swing surface \eqref{swgsurf} is homologous to the interval $\cal A$ at the boundary and its area in Planck units reproduces the entanglement entropy of BMSFTs
  \eq{
  S_{\cal A} = \frac{\mathrm{Area}(\gamma_{\cal A}) }{4 G} = \frac{l_u }{2G l_\zc}. \label{HEEpoincare}
  }
%


\subsubsection{The first law in flat$_3$/BMSFT} \label{se:flatfirstlaw}

We now show that the variation of the modular Hamiltonian matches the infinitesimal gravitational charge associated with the modular flow generator~\eqref{bulkmodflow} and derive the first law of entanglement entropy. This provides an explicit example of the general discussion in Section~\ref{se:generalfirstlaw} in the context of the flat$_3$/BMSFT correspondence. 

The gravitational charge defined in \eqref{chargedef} is given by an integral over the $(d-1)$-form ${\bm \chi_\xi[g,\delta g]}$ which, in Einstein gravity, is given by 
  \eq{
   {\bm \chi}_{\xi} [g, \delta g]   =  \frac{{\bm \epsilon}_{\mu\nu}}{16\pi G}  \Big( \xi^\mu \nabla_\sigma \delta g^{\nu\sigma} -\xi^\mu\nabla^\nu \delta g_\sigma^{\; \sigma}+\xi_\sigma\nabla^\nu \delta g^{\mu\sigma} +\frac12  \delta g_\sigma^{\; \sigma} \nabla^\nu \xi^\mu - \delta g^{\sigma \nu}\nabla_\sigma \xi^\mu \Big)
  \label{chidef},}  
where ${\bm \epsilon_{\mu\nu}} =\frac{1}{(d-1) !} \epsilon_{\mu\nu \mu_3 \dots \mu_{d+1}} dx^{\mu_3}\wedge \dots \wedge dx^{\mu_{d+1}}$. For asymptotically flat spacetimes satisfying the boundary conditions~\eqref{flatbc}, the on-shell fluctuations of the background metric \eqref{huuphi} are given by
\be
   \delta g_{\mu\nu} dx^\mu dx^\nu = {\delta \Theta}(\zc) du^2 + \Big[ {\delta \Xi}(\zc) + \frac{1 }{ 2} u \partial_{\zc} {\delta \Theta}(\zc) \Big] du d\zc, \label{deltag}
\ee
where $\delta \Theta(\zc)$ and $\delta \Xi(\zc)$ are two arbitrary functions. For these on-shell fluctuations, the components of the one-form ${\bm \chi_\xi[g,\delta g]}= \chi_u du+\chi_r dr +\chi_\zc d\zc$ read
\eq{
\begin{split}
\chi_u  &= -\frac{-2 r Y \delta \Theta  + T  \delta\Theta' + 2 Y' \big( \delta\Xi +u \delta\Theta ' \big) }{32 \pi G r}, \qquad \chi_r =  \frac{\big(T + u Y' \big) \big(2 \delta\Xi + u \delta\Theta' \big)}{32\pi G r^2}, \\
\chi_\zc&= \frac{1}{16\pi G} \bigg\{T \delta\Theta  + 2 Y \delta\Xi + \frac{1}{r} \partial_\zc \Big[r u Y \delta\Theta - \frac{1}{2} \big(T +u Y' \big) \big(2 \delta\Xi  + u \delta\Theta'\big) \Big] \bigg\},
\end{split}\label{chiurphi}
}
where we have suppressed the $\zc$-dependence of $\delta \Theta$, $\delta \Xi$, $T$, and $Y$ for convenience. 

As discussed in section~\ref{se:generalfirstlaw}, the gravitational charge $\delta {\cal Q}_\xi^{\cal C}[g] = \int_{\cal C}{\bm \chi}_{\xi} [g, \delta g] $ acquires a different physical meaning depending on the choice of the curve $\cal C$. In particular, when the charge is evaluated along the boundary interval $\cal A$ we obtain
\be\label{chargeonhorizon}
\delta \mathcal Q^{\mA}_\xi[ g]  =  \frac{1}{ 16\pi G}\int_{\zc_-}^{ \zc_+}    \big[ T(\zc) {\delta \Theta} (\zc) +2 Y(\zc) {\delta \Xi} (\zc)  \big] d\zc,
\ee
where we used integration by parts and we note that the boundary term vanishes exactly at the endpoints of the interval. Using  the holographic dictionary between the BMSFT currents $\mathcal P(\zc)$, $\mathcal J(\zc)$ and the bulk data $\Theta(\zc)$, $\Xi(\zc)$ given in \eqref{dictionary}, the gravitational charge~\eqref{chargeonhorizon} matches the modular Hamiltonian \eqref{ModHam} derived on the field theory side of the correspondence,
\be
\delta \mathcal Q^\mA_\xi[g]  =\delta \EV{\mathcal  H_{\zeta}  }=\delta \EV{\mathcal  H_{mod}  },
\ee
verifying the holographic dictionary~\eqref{s2:dHQb}.

On the other hand, as shown explicitly in~\cite{Apolo:2020bld}, the gravitational charge evaluated on the swing surface $\gamma_\mA$ is holographically dual to the entanglement entropy~\eqref{s2:dsQA}. Since $\gamma_\mA$ is homologous to the interval $\mA$ at the boundary, Stoke's theorem guarantees that $\delta {\cal Q}_\xi^{\gamma_{\mA}}[g] = \delta {\cal Q}_\xi^{\mA}[g]$. These results allow us to establish the first law of entanglement entropy in flat$_3$/BMSFT, namely
 \eq{
  \delta S_{\cal A}= \delta {\cal Q}_\xi^{\gamma_{\mA}}[g]=\delta {\cal Q}_\xi^{\mA}[g]= \delta \langle{\cal H}_{mod} \rangle. \label{firstlaw}
  }
  %


\subsection{The modular Hamiltonian and zero-mode backgrounds} \label{se:zeroMode}

In this section we generalize the results of previous sections to more general states in the bulk and boundary sides of the flat$_3$/BMSFT correspondence. More concretely, we derive the modular flow generator and the modular Hamiltonian for thermal states in BMSFTs which can be obtained by placing the theory on a thermal cylinder satisfying the following identification of coordinates
\eq{
(u, \phi) \sim (u, \phi + 2\pi) \sim (u + i \beta_u, \phi - i \beta_\phi), \label{thermalcyl}
}
where $\beta_u$ and $\beta_\phi$ are the thermodynamic potentials. In the bulk, these states are dual to flat cosmological solutions described by the zero-mode backgrounds
\eq{
ds^2 =M du^2 -2 dudr +J du d\phi +r^2 d\phi^2, \qquad M> 0, \label{zeroMode}
}
where the mass $M$ and angular momentum $J$ are related to the thermodynamic potentials $\beta_u$ and $\beta_\phi$ by
  \eq{
  \beta_u = \frac{\pi J}{M^{3/2}}, \qquad \beta_\phi = \frac{2\pi}{\sqrt{M}}.
  }
Although we assume that $M > 0$ throughout this section, our results also hold for the $M < 0$ case after analytic continuation.

Let us parametrize the interval $\mA$ at the boundary in terms of the endpoints
\eq{
\p \mA = \{ (u_-,\phi_-),(u_+,\phi_+)\}, \qquad l_u \equiv u_+ - u_-, \quad l_\phi \equiv \phi_+ - \phi_-. \label{interval1}
}
The modular flow generator can be obtained using various methods: from the generalized Rindler method, via a BMS transformation mapping the modular flow generator \eqref{modularflow} from the plane  to the thermal cylinder \eqref{thermalcyl}, or by repeating the calculation in Section~\ref{flatse:modflow}, but with the vacuum symmetry generators on the plane \eqref{BMSgenerator} replaced by the generators on the thermal cylinder. The resulting modular flow generator can be written as
\eq{
 \zeta &=\big[ T(\phi) +u Y'(\phi)  \big] \p_u +Y(\phi) \p_\phi, \label{modularflowMJ}
}
where the $T(\phi)$ and $Y(\phi)$ functions are given by 
  \eq{
 T (\phi) &= \frac{\pi }{ 2 M \sinh\Big(\tfrac{\sqrt{M} l_{\phi}}{2}\Big)} \bigg\{  (J l_\phi + 2 l_u M) \bigg[ \mathrm{coth}\Big(\tfrac{\sqrt{M} l_{\phi}}{2}\Big) \cosh(\sqrt{M} \Delta\phi)- \mathrm{csch}\Big(\tfrac{\sqrt{M} l_{\phi}}{2}\Big)   \bigg]   \notag \\ 
  &  \hspace{.45cm} + \frac{2 J }{ \sqrt{M}} \bigg[ \cosh(\sqrt{M} \Delta\phi) -  \cosh \Big(\tfrac{\sqrt{M} l_{\phi}}{2}\Big)\bigg]- 2J \Delta\phi \sinh(\sqrt{M}\Delta\phi) \bigg\}, \label{Ttilde} \\
  \! Y (\phi) &= \frac{2\pi }{ \sqrt{M}\sinh\Big(\tfrac{\sqrt{M} l_{\phi}}{2}\Big)}\Big[ \cosh\Big(\tfrac{\sqrt{M} l_{\phi}}{2}\Big) - \cosh(\sqrt{M} \Delta\phi)\Big], \label{Ytilde}
  }
with $\Delta \phi \equiv \phi - (\phi_+ + \phi_-)/2$. For convenience, we have written eqs.~\eqref{Ttilde} and~\eqref{Ytilde} directly in terms of the zero-mode charges $M$ and $J$ instead of the thermodynamic potentials $\beta_u$ and $\beta_\phi$. A similar calculation to the one described in Section \ref{flatse:modH} then shows that, up to a constant, the modular Hamiltonian for a single interval on a thermal state is given by
  \eq{
{\cal H}_{\zeta}= -\frac{1}{2\pi} \int_{\phi_-}^{\phi_+} d\phi  \big[ T(\phi)\mathcal P(\phi)+ Y(\phi) \mathcal J(\phi) \big]. \label{ModHamMJ}
}

Let us now consider the bulk side of the flat$_3$/BMSFT correspondence. The modular flow generator \eqref{modularflowMJ} can be extended into an exact Killing vector in the bulk whose components are given in terms of the $T(\phi)$ and $Y(\phi)$ functions by
 \eq{
 \xi^u &=  T(\phi) + u  Y'(\phi), \quad  \xi^\phi = Y(\phi) -\frac{u}{r} Y''(\phi)-\frac{1}{r} T'(\phi), \quad \xi^r =-\frac{J}{2r} \p_\phi \xi^u -r \partial_\phi \xi^\phi. \label{xiMJ}
}
The gravitational charge associated with the bulk modular flow generator $\xi$ evaluated on the boundary interval $\mA$ is then given by
\eq{
\delta \mathcal Q^{\mA}_\xi[ g] =\int_{\cal A} {\bm \chi}_{\xi} [ g,\delta g] = \frac{1}{ 16\pi G}\int_{\phi_-}^{\phi_+}   \Big[T(\phi) {\delta\Theta} (\phi) +2 Y(\phi) {\delta\Xi} (\phi)    \Big] d\phi. \label{bulkModMJ}
}
Using the holographic dictionary \eqref{dictionary}, we find that the infinitesimal gravitational charge \eqref{bulkModMJ} agrees with the variation of the modular Hamiltonian, namely
  \eq{
  \delta \mathcal Q^{\mA}_\xi[g] =  \delta \langle{\cal H}_{\zeta}\rangle =  \delta \langle{\cal H}_{mod}\rangle.
  }
  This provides additional evidence for the identification \eqref{s2:dHQb} discussed in Section~\ref{se:generalfirstlaw}.
  
 Finally,  as shown in~\cite{Apolo:2020bld}, the infinitesimal gravitational charge evaluated along the swing surface equals the variation of the entanglement entropy, namely  $ \delta {\cal Q}_\xi^{\gamma_{\cal A}}[g] = \delta S_{\cal A}$. Thus, using Stoke's theorem, we can generalize the first law of entanglement entropy to thermal states in BMSFTs.


\section{(W)AdS$_3$/WCFT} \label{se:wcft}

In this section we consider another class of holographic dualities where warped conformal symmetries play an essential role. On the one hand, we have warped conformal field theories (WCFTs) --- two dimensional field theories featuring a Virasoro-Kac-Moody algebra~\cite{Hofman:2011zj, Detournay:2012pc}. Explicit models of WCFTs include chiral Liouville theory~\cite{Compere:2013aya}, Weyl fermions~\cite{Hofman:2014loa}, and scalars~\cite{Jensen:2017tnb}. On the other hand, Virasoro-Kac-Moody algebras arise as the asymptotic symmetry algebras of a variety of models that admit warped AdS$_3$ spacetimes~\cite{Anninos:2008fx, Compere:2009zj}, as well as AdS$_3$ spacetimes in Einstein gravity with Dirichlet-Newman boundary conditions~\cite{Compere:2013bya}. We refer to this class of holographic models as (W)AdS$_3$/WCFT. Recent developments in (W)AdS$_3$/WCFT include reproducing the Bekenstein-Hawking entropy of black holes from an analog of the Cardy formula in WCFT~\cite{Detournay:2012pc}, average OPE coefficients of heavy-heavy-light correlators~\cite{Song:2019txa}, bulk and boundary studies of entanglement entropy~\cite{Castro:2015csg,Song:2016pwx,Song:2016gtd,Azeyanagi:2018har,Wen:2018mev, Gao:2019vcc}, one-loop partition functions~\cite{Castro:2015uaa}, correlation functions~\cite{Song:2017czq}, the modular bootstrap~\cite{Apolo:2018eky}, quantum chaos~\cite{Apolo:2018oqv}, R\'enyi mutual information~\cite{Chen:2019xpb}, bulk reconstruction~\cite{Lin:2019dji}, complexity~\cite{Ghodrati:2019bzz}, connections to complex SYK models~\cite{Chaturvedi:2018uov,Chaturvedi:2020abc}, as well as the emergence of warped conformal symmetries in non-extremal Kerr black holes~\cite{Aggarwal:2019iay}.

In this paper we will further develop our understanding of holographic entanglement entropy in (W)AdS$_3$/WCFT by studying in detail the modular Hamiltonian and deriving the first law of entanglement entropy from both field theory and holography. In what follows, we will first consider the holographic correspondence between AdS$_3$ gravity with Dirichlet-Neumann boundary conditions~\cite{Compere:2013bya} and two-dimensional warped conformal field theories~\cite{Detournay:2012pc}. We will generalize our results to warped AdS$_3$ backgrounds in Section~\ref{se:wads}.


\subsection{A brief introduction to AdS$_3$/WCFT} \label{wcftsuse:adswcft}
In this section we briefly review three-dimensional gravity with CSS boundary conditions~\cite{Compere:2013bya} and two-dimensional warped conformal field theories~\cite{Detournay:2012pc}. 

\subsubsection{AdS$_3$ with CSS boundary conditions}

Three-dimensional gravity with a negative cosmological constant admits consistent sets of Dirichlet and Dirichlet-Neumann boundary conditions at the conformal boundary. Dirichlet boundary conditions correspond to the standard boundary conditions of Brown and Henneaux where the boundary metric is nondynamical~\cite{Brown:1986nw}. These boundary conditions play an important role in the AdS/CFT correspondence as they lead to an asymptotic symmetry group described by two copies of the Virasoro algebra, matching the symmetries of the dual CFT. Alternatively, Dirichlet-Neumann boundary conditions render some components of the boundary metric dynamical and lead to different sets of asymptotic symmetries at the boundary~\cite{Compere:2013bya,Troessaert:2013fma,Avery:2013dja,Apolo:2014tua,Apolo:2015fja}.

In this paper we are interested in the CSS boundary conditions~\cite{Compere:2013bya} where one of the components of the metric becomes dynamical. Denoting the coordinates of AdS$_3$ by $u = \varphi + t$, $v = \varphi - t$, and $r$, the CSS boundary conditions are given by
  \eq{
  \begin{gathered}
 \! g_{rr} = \frac{1}{r^2} + {\cal O}(1/r^{4}), \qquad g_{ru} =  {\cal O}(1/r^{3}), \qquad g_{rv} =  {\cal O}(1/r^{3}), \qquad g_{uv} = \frac{r^2}{2} + {\cal O}(1), \\
  g_{uu} =  r^2  J'(u) +  {\cal O}(1), \qquad g_{vv} = \frac{6\Delta}{c} +  {\cal O}(1/r),
  \end{gathered}\label{wcft:cssbc}
  }
where $c =3 \ell/2G$, $G$ is Newton's constant, we have set the AdS scale $\ell$ to one, and $J(u)$ is an arbitrary function. These boundary conditions depart from those of Brown and Henneaux in the second line of \eqref{wcft:cssbc}, which tells us that ($i$) the leading component of $g_{uu}$ (the boundary metric) is dynamical; while ($ii$) the subleading component of $g_{vv}$ (the expectation value of the stress tensor in Brown-Henneaux) is a constant denoted by $\Delta$.  Finally, we note that the boundary conditions \eqref{wcft:cssbc} are compatible with varying values of $\Delta$ which we assume to be the case throughout this paper (see Appendix B of~\cite{Compere:2013bya}).  

In the Fefferman-Graham gauge, \eqref{wcft:cssbc} is sufficient to determine the most general solution to the equations of motion of pure Einstein gravity which is parametrized by $\Delta$ and two chiral functions $L(u)$ and $J(u)$,
\eq{
\begin{split}
    ds^{2} &=  \frac{dr^{2} }{ r^{2}} + r^{2} du \big[ dv + J'(u) du \big ] +  {  L} (u) (du)^{2} +  \Delta \big [ dv +  J'(u) du \big ]^{2} \\
    & \hspace{.5cm} + \frac{1 }{ r^{2}}  \Delta   {  L} (u) du \Big [ dv +  J'(u) du \Big ].  \label{wcft:cssfg}
  \end{split}
}
When $J(u)$ is a constant, \eqref{wcft:cssfg} describes a subset of the space of solutions of three-dimensional gravity with Brown-Henneaux boundary conditions. In particular, this means that the AdS$_3$ vacuum with $ {L} (u) = \Delta = -1/4$, as well as BTZ black holes at temperatures $T_R =\frac{1}{\pi} \sqrt{\Delta}$ and $T_L =  \frac{1}{\pi} \sqrt{ {L} }$ where $ {L} (u) \to  {L} $ is a constant, are also solutions of three-dimensional gravity with CSS boundary conditions. When $L(u)$ or $J'(u)$ are not constants, we interpret the corresponding backgrounds as solutions dressed with additional boundary gravitons.

The boundary conditions~\eqref{wcft:cssbc} admit an asymptotic symmetry group described by a Virasoro-Kac-Moody algebra. These symmetries are generated by the state-dependent asymptotic Killing vector~\cite{Compere:2013bya}
   \eq{
  \eta = \epsilon(u) (\p_u + \p_v) + \frac{\sigma(u)}{2 \sqrt{-c \Delta/ 6k}} \p_v - \frac{r \epsilon'(u)}{2} \p_r + {\cal O}(1/r), \label{wcft:asymptotickillings}
  }
where $k$ is the level of the $U(1)$ algebra.\footnote{A state-independent version of the asymptotic Killing vectors~\eqref{wcft:asymptotickillings} also exists~\cite{Compere:2013bya}, but integrability of the corresponding charges requires $\Delta$ to be a fixed constant.} Expanding the Killing vectors in terms of Fourier modes via 
$\epsilon(u) = \Sigma_n \epsilon_n e^{i n u}$ and $\sigma(u) = \Sigma_n \sigma_n e^{i n u}$ one finds the following conserved charges
  \eq{
 {\cal Q}_{\epsilon_n} &= \frac{c}{12\pi} \int d\varphi \,e^{i n u} \Big \{  {L} (u) - \Delta\,\big [1 +  J'(u) \big]^2 \Big\},  \label{wcft:Ln}\\
 {\cal Q}_{\sigma_n} &=  \frac{1}{2\pi} \int d\varphi\, e^{i n u}  k \sqrt{-c \Delta /6k}  \big [ 1 + J'(u) \big ],\label{wcft:Pn}
  }
which satisfy a canonical Virasoro-Kac-Moody algebra with central charge $c$ and level $k$,
  \eqsp{
  [{\cal Q}_{\epsilon_m}, {\cal Q}_{\epsilon_n}] &= (m - n) {{\cal Q}_{\epsilon_{m+n}}} + \frac{c}{12} m^3 \delta_{m+n,0}, \\
  [{\cal Q}_{\epsilon_m}, {\cal Q}_{\sigma_n}] &= - n {\cal Q}_{\sigma_{m+n}},\\
  [{\cal Q}_{\sigma_m}, {\cal Q}_{\sigma_n}] &= \frac{k}{2} m\, \delta_{m+n,0}. \label{wcft:virkacmoody}
  }
When the level is negative, the $U(1)$ charges~\eqref{wcft:Pn} are real for BTZ black holes ($\Delta > 0$) and imaginary otherwise ($\Delta < 0$). This property of the $U(1)$ charges is a direct consequence of the state-dependent normalization of the Killing vector \eqref{wcft:asymptotickillings}, the latter of which is necessary to match the symmetries in the bulk and boundary sides of the AdS$_3$/WCFT correspondence. In particular, a negative level implies that the $U(1)$ charge of the vacuum is imaginary. This feature plays an important role in the holographic description of 3d gravity with CSS boundary conditions as it guarantees, for example, that the black hole and entanglement entropies computed in the field theory are both real quantities~\cite{Detournay:2012pc,Castro:2015csg,Song:2016gtd}. In this paper we will assume that the $U(1)$ level of the Virasoro-Kac-Moody algebra is negative (this means that we can set $k = -1$ but we keep $k$ intact in all our calculations).


\subsubsection{Warped CFT} \label{wcftsuse:wcft}

Three-dimensional gravity with CSS boundary conditions is conjectured to be dual to a two dimensional warped CFT on the boundary. The latter is a nonrelativistic quantum field theory with enhanced local symmetries described by a chiral Virasoro-Kac-Moody algebra~\cite{Hofman:2011zj}. In contrast to (chiral) CFTs with internal $U(1)$ symmetries\footnote{The relationships between chiral CFTs and warped CFTs is discussed in detail in Section~2 of~\cite{Chaturvedi:2018uov}.}, both the Virasoro and $U(1)$ parts of the WCFT algebra correspond to spacetime symmetries generated by the coordinate transformations
  \eq{
  x' = f(x), \qquad y' =  y + g(x), \label{wcft:fg}
  }
where $(x,y)$ denote the coordinates of the WCFT and $f(x)$, $g(x)$ are two arbitrary functions. 

The warped conformal transformation~\eqref{wcft:fg} is generated by two chiral currents denoted by $T(x)$ and $P(x)$ which transform as 
  \eq{
  T'(x{}') & = \Big ( \frac{\p x}{\p x{}'} \Big)^2 \Big [ T(x) - \frac{c}{12} \{ x{}', x \} \Big ] + \frac{\p x}{\p x{}'} \frac{\p y}{\p x{}'} P(x) - \frac{k}{4} \Big ( \frac{\p y}{\p x{}'} \Big )^2, \label{wcft:Tprime} \\
  P'(x{}') & = \Big ( \frac{\p x}{\p x{}'} \Big ) \Big [ P(x) + \frac{k}{2} \frac{\p y{}'}{\p x} \Big ], \label{wcft:Pprime}
  }
where $\{\cdot,\cdot\}$ denotes the Schwarzian derivative. When $g(x)$ is constant,  $T(x)$ and $P(x)$ transform as the holomorphic components of the stress-tensor and a $U(1)$ current, respectively. Furthermore, we see that the $T(x)$ and $P(x)$ currents transform anomalously under warped conformal transformations, as indicated by the central charge $c$ and $U(1)$ level $k$. In particular, when the theory is defined on the {\it canonical} cylinder where
  \eq{
  (x,y) \sim (x + 2\pi, y),\label{wcft:canonicalcircle}
  }
   the generators of warped conformal transformations are given by
   \eq{
  \mL_n &= - \frac{1}{2\pi} \int dx \, e^{inx} {T}(x), \qquad \mP_n =- \frac{1}{2\pi} \int dx\, e^{inx} {P}(x), \label{wcft:wcftcharges}
  }
and satisfy a Virasoro-Kac-Moody algebra with central charge $c$ and level $k$. 

When describing WCFTs it is important to account for spectral flow transformations that leave the Virasoro-Kac-Moody algebra unchanged but shift the vacuum expectation values of the zero-mode charges. On the canonical cylinder \eqref{wcft:canonicalcircle}, a WCFT is characterized by vacuum charges that are parametrized in terms $c$, $k$, and a variable $\mu$ by
\eq{ 
{\mL}_0^{vac} \equiv \Vev{\mL_0} = -\frac{c}{24}-\frac{\mu^2 k}{4} , \qquad \mP_0^{vac} \equiv \Vev{\mP_0}= -\frac{i\mu k}{2}. \label{wcft:vacuumcharges}
}
A spectral flow transformation 
\eq{
  x = x', \qquad y = y' - i \mu x,
  }
maps the canonical cylinder to the \emph{reference} cylinder $(x', y')$ where the theory features the following zero-mode charges and a twisted identification of coordinates
  \eq{
   \mL_0'^{\,vac}  = -\frac{c}{24}, \quad \mP_0'^{\,vac} = 0, \qquad (x', y') \sim (x' + 2\pi, y' - 2\pi i \mu).  \label{wcft:referenceplane}
  }
In addition, note that the reference/canonical cylinder is related to the reference/canonical plane by a chiral exponential map $z = e^{ix}$ without any spectral flow involved.

Let us now describe the holographic dictionary relating the bulk and boundary sides of the AdS$_3$/WCFT correspondence. We first note that \eqref{wcft:fg} is the same symmetry transformation induced at the boundary by the asymptotic Killing vectors~\eqref{wcft:asymptotickillings} provided that we identify the bulk $(u,v)$ and boundary $(x,y)$ coordinates via the state-dependent map
  \eq{
  u = x, \qquad v  = x + \frac{k y}{2 {\cal Q}_{\sigma_0}}.  \label{wcft:map}
  }
Correspondingly, the gravitational charges~\eqref{wcft:Ln} and~\eqref{wcft:Pn} in the bulk are mapped to the Virasoro-Kac-Moody generators at the boundary, namely ${\cal Q}_{\epsilon_n} \to {\mL}_n$ and ${\cal Q}_{\sigma_n} \to {\mP}_n$. In particular, the bulk $\{ L (u),  J'(u), \Delta\}$ and boundary $\{ {T}(x), {P}(x)\}$ variables are related by
  \eq{
  {T}(x) &= -\frac{c}{6}\Big\{  L (u) - \Delta\,\big [1 +  J'(u) \big]^2 \Big\}, \qquad {P}(x) = - k \sqrt{-c\Delta/6k}  \big [ 1 + J'(u) \big ]. \label{wcft:currentdictionary}
  }
Finally, by comparing the vacuum charges~\eqref{wcft:wcftcharges} to the corresponding values of the gravitational charges for the global AdS$_3$ vacuum, we find that the $\mu$ parameter is given by
  \eq{
  \mu = \sqrt{-\frac{c}{6k}}.
  }
  %


\subsection{The boundary modular Hamiltonian} \label{wcftse:boundary}

In this section we derive the modular Hamiltonian associated with a single interval in a WCFT. We pay special attention to the contributions from anomalies as the latter are necessary to match variations of the modular Hamiltonian in the bulk and boundary sides of the AdS$_3$/WCFT correspondence. Finally, we compute the variation of the entanglement entropy resulting from perturbations of the vacuum and show that the first law of entanglement holds in WCFTs.

\subsubsection{Modular flow for a single interval} \label{wcftse:modHalgebra}

We begin by deriving general expressions for the modular flow generator $\zeta$ associated with a single interval ${\cal A}$ in WCFTs on the plane and the (thermal) cylinder. As discussed in Section~\ref{se:modularH}, the modular flow generates a symmetry transformation in the causal domain of dependence of subregion ${\cal A}$. For special states in WCFTs including the vacuum and thermal states, the modular flow generator admits a geometric description as a linear combination of the $SL(2,R)\times U(1)$ generators that leave the vacuum invariant. We first derive the modular flow generator on the reference plane following the requirements listed in Section~\ref{se:modularH} and then use a warped conformal transformation to map this result to the (thermal) cylinder.

\begin{figure}[!h]
  \centering
  \includegraphics[scale=0.3333]{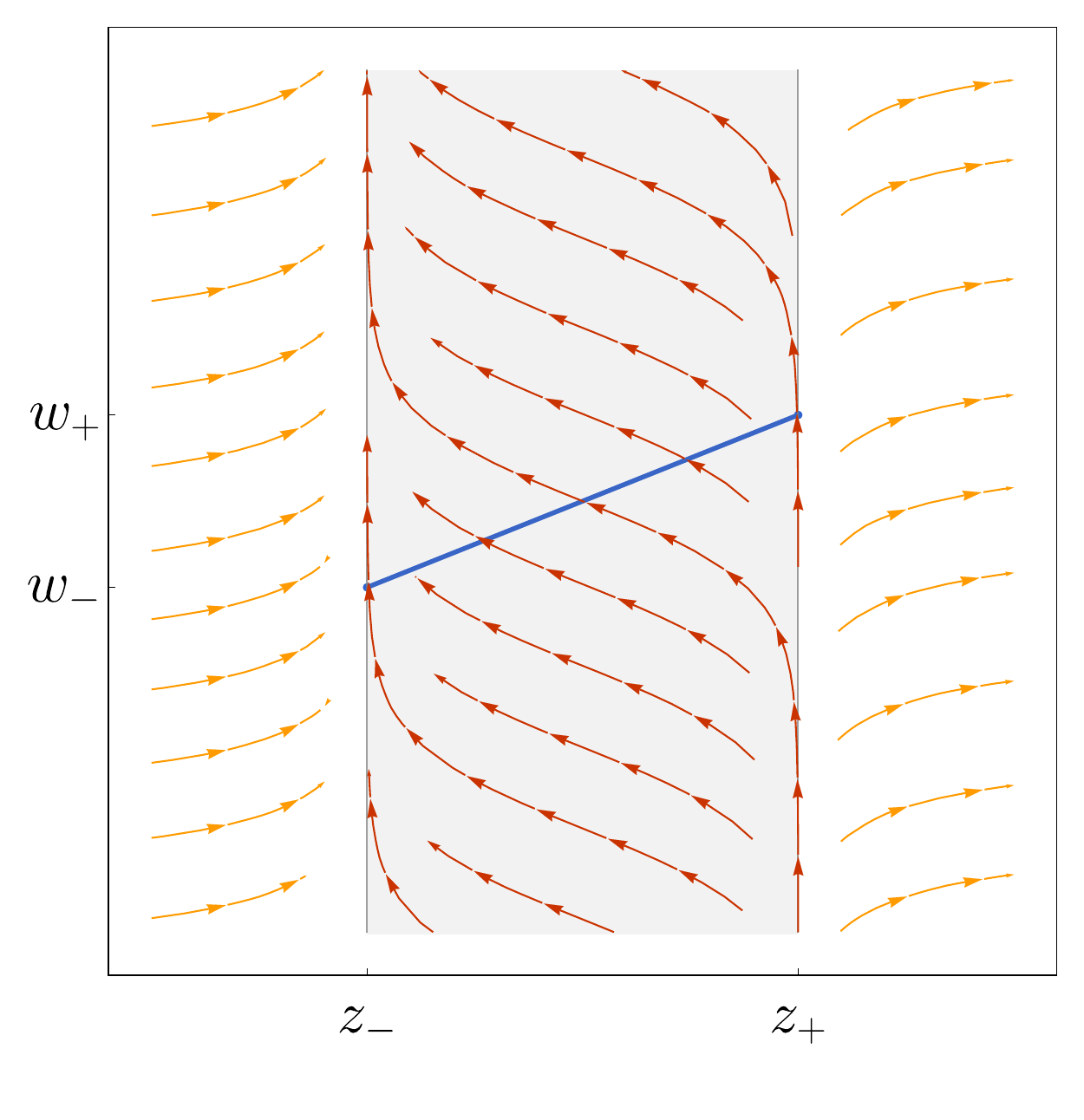}
  \caption{The modular flow of a WCFT on the plane with $\mu \ne 0$. The interval ${\cal A}$ (blue line) lies between $(z_-, w_-)$ and $(z_+, w_+)$ and its causal domain $\cal D$ is shown in gray. The modular flow \eqref{wcft:ktplane} in $\cal D$ is shown in red while the modular flow in the complement is shown in orange.}
  \label{wcftfig:modularflow}
  \end{figure}

Let us parametrize the entanglement interval ${\cal A}$ by the endpoints
  \eq{
  \p \mA=\{ (z_-,w_-),(z_+,w_+)\}, \qquad l_z \equiv z_+ - z_-, \quad l_w \equiv w_+ - w_-,    \label{wcft:interval2}
	}
where $(z,w)$ denote the coordinates on the reference plane where the zero-mode charges vanish and $(z, w) \sim (z, w - 2\pi i \mu)$. In analogy with BMSFTs, the causal domain of dependence ${\cal D}$ of WCFTs consists of a vertical strip along $w$ which is bounded by $z_-$ and $z_+$ (see Fig.~\ref{wcftfig:modularflow})~\cite{Castro:2015csg}. Following the strategy outlined in Section~\ref{se:modularH}, we are looking for a linear combination of global $SL(2,R) \times U(1)$ generators that leaves the causal domain of ${\cal A}$ intact. On the plane, the global symmetry generators are given by
  \eq{
  \ell_1 = - z^2 \p_z, \qquad \ell_0 = - z \p_z, \qquad \ell_{-1} = - \p_z, \qquad \bar{\ell}_0 = -\p_w, \label{wcft:generators}
  }
such that the modular flow generator $\zeta$ can be written as
  \eq{
  \zeta = \sum_{i=-1}^{1} a_i \ell_i + \bar{a}_0 \bar{\ell}_0. \label{wcft:zetaparametrization}
  }
By demanding that $\zeta$ vanishes along the boundaries $\partial {\cal D}$ of the causal domain, i.e.~that $\zeta|_{\partial {\cal A}} \propto \partial_w$, the $a_i$ coefficients are found to satisfy
  \eq{
  a_1 = c_0 \frac{1}{z_+ - z_-}, \qquad a_0 = -c_0  \frac{z_+ + z_-}{z_+ - z_-}, \qquad a_{-1} = c_0 \frac{z_+ z_-}{z_+ - z_-},
  }
where $c_0$ is an undetermined overall constant. This residual degree of freedom is fixed by requiring that $e^{i\zeta/2}$ maps a point in ${\cal D}$ to its complement or, alternatively, that $e^{i\zeta}$ maps a point in ${\cal D}$ back to itself. Let us consider a point with coordinates $(z(0),w(0))=(\frac{z_- e^{d_0 z_+} - z_+ e^{ d_0 z_-}}{e^{d_0 z_+} - e^{ d_0 z_-}},w)$ where $d_0$ is a real parameter. Under the flow $e^{s \zeta}$, the point $(z(0), w(0))$ becomes
\eq{
	(z(s),\, w(s))=\left( \frac{z_- e^{d_0 z_+} - z_+ e^{c_0 s} e^{ d_0 z_-}}{e^{d_0 z_+}-e^{c_0 s} e^{ d_0 z_-}},
	w - \bar{a}_0 s \right).
}
Requiring that $z(0)=z(i)$ we obtain
\eq{
	c_0 = 2\pi, \label{wcft:c0value}
}
such that, up to a sign, the $a_i$ coefficients satisfy
  \eq{
  a_1 = 2\pi \frac{1}{z_+ - z_-}, \qquad a_0 = -2\pi  \frac{z_+ + z_-}{z_+ - z_-}, \qquad a_{-1} = 2 \pi \frac{z_+ z_-}{z_+ - z_-} \label{wcft:a_i}.
  }
Note that these parameters agree with the holomorphic coefficients of the modular flow generator in 2d CFTs given in~\cite{Czech:2019vih}. 

On the other hand, the parameter $\bar{a}_0$ is fixed as follows. As observed in Section 3.1 of~\cite{Chen:2019xpb}, the replica symmetry in WCFTs is implemented on the reference plane with identification $(z, w) \sim (z e^{2\pi i}, w - 2\pi i\mu)$. This means that, as we map a point in $\cal D$ back to itself using the modular flow, we must also translate the $w$ coordinate an amount $2\pi\mu$.  Consequently, we find that $\bar{a}_0$ satisfies
  \eq{
  \bar{a}_0 = 2\pi\mu.
  }
Using the $SL(2,R) \times U(1)$ generators \eqref{wcft:generators}, the modular flow generator on the causal domain $\cal D$ is given on the reference plane by
  \eq{
  \zeta &= 2\pi \mu \bar{\ell}_0 + \frac{2\pi}{z_+ - z_-} \big [ \ell_1 - (z_+ + z_-) \ell_0 + z_+ z_- \ell_{-1} \big], \\
   &=-2\pi\mu \p_w - \frac{2 \pi}{z_+ -z_-} \big [z_+ z_- - (z_+ + z_-) z + z^2 \big] \p_z. \label{wcft:ktplane}
  }
The modular flow generated by~\eqref{wcft:ktplane} is illustrated in Fig.~\ref{wcftfig:modularflow}. 

From the modular flow generator on the reference plane \eqref{wcft:ktplane}, we can obtain the corresponding expression on the thermal cylinder with identification
  \eq{
  (x, y) \sim (x + i \beta, y - i\theta), \label{wcft:thermalcircle}
  }
by the following warped conformal transformation~\cite{Song:2016gtd}
  \eq{
  z = e^{\frac{2\pi x}{\beta}}, \qquad w = y + \bigg(\frac{\theta - 2\pi\mu}{\beta}\bigg) x.\label{wcft:thermalcc}
  }
Using \eqref{wcft:thermalcc} we find that the modular flow generator on the thermal cylinder is given by
  \eq{
  \!\!\! \zeta = -2\pi\mu \p_y  - \beta \,  \Bigg[ \frac{ \cosh\frac{\pi(2x - x_+ - x_-)}{\beta}  - \cosh\frac{\pi l_x}{\beta} }{\sinh \frac{\pi l_x}{\beta}}  \Bigg] \bigg( \partial_x - \frac{\theta - 2\pi\mu}{\beta} \partial_y \bigg), \label{wcft:kt3}
  }
where $x_+$ and $x_-$ denote the boundaries of ${\cal D}$ in the $(x,y)$ coordinates and $l_x \equiv x_+ - x_-$. 
  

\subsubsection{Modular chaos}  \label{wcftse:modularalgebra}
 
We now consider the bound on modular chaos discussed in Section~\ref{se:modularalgebra} and show that this bound is saturated in WCFTs for perturbations of the modular flow generator that result from deformations of the shape of the interval $\mA$. 

The modular flow generator  for single intervals on the vacuum state is given by a linear combination of the vacuum symmetry generators \eqref{wcft:ktplane}. Consequently, the variation of the modular flow generator  due to shape deformations can be written as
\eq{
\delta \zeta=\delta_+\zeta +\delta_-\zeta, \qquad \delta_\pm\zeta \equiv \sum_{i=-1}^1 \delta z_\pm ({\p_{z_\pm}}a_i) \ell_i  ,
}
where the $a_i$ coefficients are given in \eqref{wcft:a_i}. One can easily verify that $\delta_\pm\zeta$ satisfies the eigenvalue problem \eqref{s2:modularchaos2} such that
 \eq{
  [\zeta, \delta_{\pm} \zeta] = \pm 2\pi  \delta_{\pm} \zeta. \label{wcft:modularalgebra}
    }
Using eq.~\eqref{wcft:kt3}, it is not difficult to show that a similar equation holds on the thermal cylinder. Geometrically, $\delta_+\zeta$ is the result of moving the endpoints of $\mA$ as follows
 \eq{
 \p{\cal A} =\{ (z_+,w_+),\, (z_-,w_-) \}\to \{ (z_++\delta z_+,w_++\delta w_+), \, (z_-,w_-+\delta w_-)\}, \label{wcft:shape}
 }
where only $z_-$ is kept fixed. We conclude that the shape deformations \eqref{wcft:shape} saturate the bound on modular chaos and correspond to modular scrambling modes \cite{deBoer:2019uem}. In terms of the causal domain $\cal D$ depicted in Fig.~\ref{wcftfig:modularflow}, these deformations move one side of the strip $\cal D$ while keeping the other side fixed. This differs from both two-dimensional CFTs, where the modular scrambling modes move the two endpoints simultaneously along a specific null direction, and from BMSFTs, where such modes move one endpoint but keep the other endpoint fixed.  Finally, as discussed in~\cite{deBoer:2019uem} and reviewed in Section~\ref{se:modularalgebra}, a bound on modular chaos has been conjectured to exist in unitary and Lorentz-invariant theories. The saturation of this bound  \eqref{wcft:modularalgebra} suggests that it is possible to extend the bound on modular chaos to WCFTs despite their lack of Lorentz invariance. 


\subsubsection{Entanglement entropy in WCFT}

We now describe the generalized Rindler transformation used in \cite{Song:2016gtd} to derive the entanglement entropy of WCFTs at finite thermodynamic potentials. In particular, we will show that the modular flow generator derived in Section~\ref{wcftse:modHalgebra} is consistent with the expression obtained from the generalized Rindler transformation. We also write down the entanglement entropy of an excited state on the thermal cylinder, an expression that will be useful in establishing the first law of entanglement entropy in WCFTs.

Let us consider a WCFT on the thermal cylinder $(x,y)\sim (x+i\beta,y-i\theta)$ where the endpoints of the interval ${\cal A}$ are parametrized by
  \eq{
    \p \mA=\{ (x_-,y_-), (x_+,y_+)\}, \qquad l_x \equiv x_+ - x_-, \quad l_y \equiv y_+ - y_-,    \label{wcft:interval}
	}
A generalized Rindler transformation is a warped conformal transformation satisfying the properties listed in Section~\ref{se:modularH} and it is given by~\cite{Castro:2015csg,Song:2016gtd}
  \eqsp{
	 \tanh \frac{\pi \tilde{x}}{\tilde{\beta}} & =\coth \frac{\pi (x_+ - x_-)}{2 \beta} \tanh \frac{\pi (2x - x_+ - x_-)}{2\beta} ,\\ 
	  \tilde{y} + \bigg( \frac{\tilde{\theta}-2\pi \mu}{\tilde{\beta}} \bigg ) \tilde{x} &= \frac{2y -y_+ -y_-}{2}  + \bigg ( \frac{\theta -2\pi \mu}{\beta}  \bigg) \bigg(\frac{2x -x_+-x_-}{2}\bigg)  , \label{wcft:rindler}
  }
where $\mu$ is the spectral flow parameter described in Section~\eqref{wcftsuse:wcft}. The Rindler transformation~\eqref{wcft:rindler} maps the causal domain of ${\cal A}$ to a generalized Rindler spacetime featuring a thermal identification $(\tilde{x},\tilde{y})\sim (\tilde{x} + i \tilde{\beta}, \tilde{y} - i \tilde{\theta})$. In particular, the modular flow generator $\zeta$ corresponds to the generator of the Rindler thermal identification \cite{Song:2017czq,Wen:2018mev} 
  \eq{
  \zeta&=\tilde{\beta}\p_{\tilde x}-\tilde{\theta}\p_{\tilde y}, \label{wcft:kt}
  }
which agrees with the general expression~\eqref{wcft:kt3} derived in Section~\ref{wcftse:modHalgebra}.

Since the Rindler transformation \eqref{wcft:rindler} is a symmetry transformation, it can be implemented as a unitary transformation between density matrices and it leaves the von Neumann entropy invariant.  As a result, the entanglement entropy of ${\cal A}$ reduces to the thermal entropy of the $(\tilde{x}, \tilde{y})$ system, the latter of which can be obtained by a generalization of Cardy's formula~\cite{Detournay:2012pc}. The entanglement entropy at finite thermodynamic potentials is then given by~\cite{Castro:2015csg,Song:2016gtd}
  \eq{
  S_{\cal A}(\beta,\theta) = - \frac{\mu k}{2} \bigg [ l_y + \bigg (\frac{\theta - 2\pi \mu}{\beta}\bigg ) l_x \bigg ] + \frac{c}{6} \log \bigg ( \frac{\beta}{\pi\varepsilon} \, \sinh \frac{\pi l_x}{\beta} \bigg ), \label{wcft:thermalee}
  }
where $\varepsilon$ is a UV cutoff.

An alternative way to calculate the entanglement entropy~\eqref{wcft:thermalee} is to use the replica trick~\cite{Castro:2015csg,Song:2017czq}. Using this approach, the entanglement entropy associated with the interval ${\cal A}$ on an excited state $\Ket{\psi}$ was computed on both the reference plane and reference cylinder in ref.~\cite{Apolo:2018oqv}.\footnote{This calculation is valid in the semiclassical limit where the central charge $c$ is large, the Sugawara combination of the conformal weight $h_{\psi}$ and $U(1)$ charge $q_{\psi}$ of $\Ket{\psi}$, namely $h_{\psi} - q_{\psi}^2/k$, scales linearly with $c$, and the vacuum block is assumed to dominate heavy-heavy-light-light four-point functions.} The change of coordinates \eqref{wcft:thermalcc} allows us to map the results of~\cite{Apolo:2018oqv} to the thermal cylinder such that the entanglement entropy $S^{\psi}_{\cal A}$ on a state $\Ket{\psi}$ with thermodynamic potentials $\beta$ and $\theta$ reads\footnote{Note that in our conventions the $U(1)$ charge $q_\psi$ in eq.~\eqref{wcft:psiee} flips sign with respect to ref.~\cite{Apolo:2018oqv}.}
  \eq{
  S^{\psi}_{\cal A} &= -\frac{\mu k}{2} \bigg [ l_y + \bigg (\frac{\theta -2\pi \mu}{\beta}\bigg ) l_x \bigg ] \! + \frac{i \mu }{\beta}q_{\psi} l_x + \frac{c}{6} \log \bigg ( \frac{{\beta \tilde{\gamma}}}{\pi  \varepsilon} \,\sinh \frac{\pi l_x}{{\beta \tilde{\gamma}}} \bigg ), \label{wcft:psiee}
  }
  where $h_{\psi}$ and $q_{\psi}$ denote the conformal weight and $U(1)$ charge of the operator ${\mathcal O}_\psi$ and $\tilde{\gamma}$ is defined by
  \eq{
  \tilde{\gamma} = \frac{1}{\sqrt{1 -\tfrac{24}{c} (h_{\psi} - q_{\psi}^2/k)}}.
  }
  %


\subsubsection{The modular Hamiltonian} \label{wcftse:modularHboundary}

The modular Hamiltonian is the conserved charge associated with the modular flow generator \eqref{wcft:kt3}. This charge is defined up to constant terms, the latter of which receive contributions from the anomalies originating from the  warped conformal transformation~\eqref{wcft:rindler}. We expect the variation of the modular Hamiltonian on a fixed subregion ${\cal A}$ to be independent of the anomaly, as the later does not depend on the state. This is indeed true in both CFTs and BMSFTs. However, additional subtleties arise in WCFTs when we try to match variations of the modular Hamiltonian with the corresponding gravitational charge in the bulk. This follows from the fact that the bulk and boundary coordinates are related by a state-dependent transformation \eqref{wcft:map}, which implies that the bulk $(l_u,l_v)$ and boundary $(l_x, l_y)$ parametrizations of the interval ${\cal A}$ satisfy
  \eq{
  l_u = l_x, \qquad l_v = l_x + \frac{k}{2 \mP_0} l_y. \label{wcft:intervalmap}
  }
For this reason we must keep track of the constant $l_y$-dependent contributions to the modular Hamiltonian.

Let us now calculate the modular Hamiltonian using the Rindler method, which allows us to keep track of the anomalies resulting from warped conformal transformations.  
We first note that the range of coordinates $(\tilde{x}, \tilde{y})$ in the generalized Rindler transformation \eqref{wcft:rindler} is infinite. Assuming that edge effects can be ignored, it is convenient to introduce a large spatial circle described by the following identification~\cite{Castro:2015csg}
  \eq{
  (\tilde{x},\tilde{y}) \sim (\tilde{x}+2 \pi a, \,\tilde{y}-2 \pi b),  \label{wcft:spatialidentification}
  }
where $a$ and $b$ are given by
\eq{
a=\frac{\tilde{\beta}}{2\pi^2}  \log \bigg (\frac{\beta}{\pi \varepsilon}\sinh \frac{l_x \pi}{\beta}\bigg),\qquad b= \bigg (\frac{\tilde{\theta} - 2\pi\mu}{\tilde \beta}\bigg )a  - \bigg (\frac{\theta - 2\pi\mu}{2\pi\beta} \bigg )l_x - \frac{l_y}{2\pi}. \label{wcft:ab}
}
This effectively puts the theory in the $(\tilde{x}, \tilde{y})$ system on a torus, which will allow us to read the modular Hamiltonian directly from the torus partition function.
Next, we introduce an additional set of coordinates $(\hat{x},\hat{y})$ which are related to the Rindler ones $(\tilde{x}, \tilde{y})$ via $(\tilde{x} ,\, \tilde{y} ) = (a \hat{x}, \hat{y}-b \hat{x})$. In these coordinates the spatial and thermal circles are respectively given by
\eq{
	(\hat{x},\hat{y}) \sim (\hat{x} + 2\pi,\hat{y}) \sim \bigg(\hat{x} + i\frac{\tilde{\beta}}{a},\hat{y} - i \tilde{\theta} + i \frac{b\tilde{\beta}}{a} \bigg).\label{wcft:canonicaltorus}
}
In particular, the spatial identification \eqref{wcft:canonicaltorus} agrees with the definition of the canonical cylinder \eqref{wcft:canonicalcircle}, which allows us to use the charges defined in \eqref{wcft:wcftcharges} as well as their vacuum expectation values \eqref{wcft:vacuumcharges}. Since the inverse temperature $\tilde\beta/a$ in the thermal circle \eqref{wcft:canonicaltorus} is small, we can use modular covariance of the partition function to approximate the partition function~\cite{Castro:2015csg,Song:2016gtd}, 
  \eq{
 Z(\hat{\beta}, \hat{\theta}) &\equiv \Tr \exp \bigg[-\frac{\tilde{\beta}}{a} \hat{\mathcal{L}}_{0} + \bigg (\tilde{\theta}-\frac{b \tilde{\beta}}{a} \bigg) \hat{\mathcal{P}}_{0}\bigg] \label{wcft:Zdef}  \\
  &\sim \exp  \bigg[  \frac{ak}{4\tilde{\beta}}  \bigg( \tilde{\theta} - \frac{b \tilde{\beta}}{a} \bigg)^2 - 2 \pi i \bigg( \frac{a\tilde{\theta}}{\tilde{\beta}} - b \bigg) \mP_0^{vac} - \frac{4\pi^2 a}{\tilde{\beta}} \mL_0^{vac} \bigg] \label{wcft:logz}\\
  &= \exp \bigg[ \frac{c}{12} \log \bigg (\frac{\beta}{\pi \varepsilon}\sinh \frac{l_x \pi}{\beta}\bigg) + {\cal O}(1/a)\bigg] , \label{wcft:constdef}
  }
  where in the third line we used eqs.~\eqref{wcft:vacuumcharges} and \eqref{wcft:ab}. The Hamiltonian associated with the generator of the thermal identification $\zeta=\hat{\beta}\p_{\hat x}-\hat{\theta}\p_{\hat y}$ can be read from \eqref{wcft:Zdef} and reads
   \eq{
  {\hat{\cal H}}_\zeta &= \frac{\tilde{\beta}}{a} \hat{\mL}_0 - \bigg(\tilde{\theta} - \frac{b \tilde{\beta}}{a} \bigg) \hat{\mP}_0 - \log Z \label{wcft:aaaa} \\
  & = - \frac{1}{2\pi} \int d\hat{x} \bigg \{ {\frac{\tilde{\beta}}{a}} \hat{T}(\hat{x}) - \bigg(\tilde{\theta}-\frac{b\tilde{\beta}}{a}\bigg) \hat{P}(\hat{x}) \bigg \} - \log Z.\label{wcft:Qktinhat}
  }
where the $\log Z$ term is given in \eqref{wcft:constdef} and is added to make the trace of the thermal density matrix $\hat{\rho} \equiv e^{- {\hat{\cal H}}_\zeta}$ equal to one. The reduced density matrix on $\cal A$ is related to $\hat \rho$ by a unitary transformation which does not affect the trace. As a result, the Hamiltonian defined in \eqref{wcft:Qktinhat} leads to the modular Hamiltonian in $\mA$ with the correct constant terms, namely $ {\cal H}_{mod}=  {\hat{\cal H}}_\zeta$.

The modular Hamiltonian in terms of the variables in the $(x,y)$ coordinates can be obtained from \eqref{wcft:Qktinhat} by performing a series of warped conformal transformations $(\hat{x}, \hat{y}) \to (\tilde{x}, \tilde{y}) \to (x,y)$. As discussed earlier, we have to pay special attention to the constant $l_y$-dependent contributions to the modular Hamiltonian as the bulk parametrization of the interval $\mA$ depends on the state. In the $(\hat{x}, \hat{y})$ coordinates, $l_y$ enters the partition function only through the combination $b/a$. When the cutoff is taken to zero, the finite $b/a$ terms in \eqref{wcft:constdef} cancel out such that $\log Z$ is independent of $l_y$.  Thus, the $l_y$-dependent contributions to the modular Hamiltonian originate from anomalies in the warped conformal transformation mapping the $(\hat x, \hat y)$ system to $(x,y)$. Keeping track of the anomalies we find that the modular Hamiltonian ${\cal H}_{mod}$ is given by
 \eq{
 {\cal H}_{mod} & =  \int_{x_-}^{x_+} \!\!dx \bigg \{ \mu P(x) + \frac{\beta}{2\pi} \bigg[ \frac{\cosh \frac{\pi (2x -x_+ -x_-)}{\beta} - \cosh \frac{\pi l_x}{\beta}}{\sinh \frac{\pi l_x}{\beta}} \bigg]\bigg [ T(x) - \bigg ( \frac{\theta - 2\pi\mu}{\beta} \bigg) P(x) \bigg] \bigg \} \notag \\
  & \phantom{==} - \frac{k \mu}{2} l_y - \bigg[ \frac{c}{12} - \frac{k}{8\pi^2}  (2\pi\mu - \theta)^2 \bigg]\bigg[ \frac{\pi l_x}{\beta} \coth \bigg(\frac{\pi l_x}{\beta}\bigg)- 1 \bigg], \label{wcft:defH}
  }
where the first line depends on the conserved currents $T(x)$ and $P(x)$ while the second line is the contribution from the anomaly and the $\log Z$ term. Note that the modular Hamiltonian is independent of the thermodynamic potentials $\tilde{\beta}$ and $\tilde{\theta}$ introduced by the generalized Rindler transformation~\eqref{wcft:rindler}. This provides a consistency check of our calculations as the tilded variables are not physical and should not appear in the final expression for ${\cal H}_{mod}$. Finally, note that there is a divergent contribution from the anomaly that cancels the divergent contribution from the $\log Z$ term in \eqref{wcft:constdef} so that the modular Hamiltonian \eqref{wcft:defH} is finite and independent of the UV cutoff. 

When $l_x$ and $l_y$ are fixed,  the second line of \eqref{wcft:defH} does not contribute to the variation of the modular Hamiltonian.  However, due to the state-dependent map~\eqref{wcft:map} characteristic of holographic WCFTs, the $l_y$-dependent term is physically relevant and necessary to match the modular Hamiltonian to the corresponding gravitational charge in the bulk. Consequently,  the variation of the modular Hamiltonian is written as    %
    \eq{
  \delta {\cal H}_{mod} &=  \int_{x_-}^{x_+} dx \bigg \{ \mu \delta P(x) + \frac{\beta}{2\pi}  \bigg[ \frac{\cosh \frac{\pi (2x -x_+ -x_-)}{\beta} - \cosh \frac{\pi l_x}{\beta}}{\sinh \frac{\pi l_x}{\beta}} \bigg] \bigg [ \delta T(x) - \bigg ( \frac{\theta-2\pi\mu}{\beta} \bigg) \delta P(x) \bigg] \notag \\
  & \phantom{=  \int_{-\frac{l_x}{2}}^{\frac{l_x}{2}} dx \bigg \{}-  \frac{\mu k \delta l_y}{2l_x}\bigg \}.  \label{wcft:deltaH} 
  }
  %
  

\subsubsection{The first law in WCFT}\label{wcftse:firstlawboundary}

The first law of entanglement entropy states that variations of the entanglement entropy equal the variations of the modular Hamiltonian, namely $\delta S_{\cal A} = \delta \langle{\cal H}_{mod}\rangle$. This is a fine-grained version of the first law of thermodynamics which allows us to check the consistency of WCFTs. We will show that the first law is indeed satisfied in these theories by studying the variation of the entanglement entropy for both zero-mode excitations as well as local excitations generated by the insertion of an operator, and comparing these results to the variation of the modular Hamiltonian derived in the previous subsection.

\subsubsection*{Zero-mode fluctuations} 

Let us first consider the change of the entanglement entropy~\eqref{wcft:thermalee} under infinitesimal variations of the zero-mode charges. The variation of the entropy $\delta S_{\mA}$ is given by
  \eq{
  \!\! \delta S_{\mA} = -\frac{\mu k}{2} \delta l_y  - \frac{\mu k}{2} l_x \delta \bigg(\frac{\theta -2\pi \mu}{\beta} \bigg) + \frac{c}{6} \bigg(1 - \frac{\pi l_x}{\beta} \coth \frac{\pi l_x}{\beta} \bigg)\frac{\delta \beta}{\beta}, 
  }
where we have included the variations of $l_y$ for completeness. In the canonical ensemble, the variations of the thermodynamic potentials are related to the variation of the thermal expectation values of the $\mL_0$ and $\mP_0$ generators. The partition function of a WCFT on a torus with canonical spatial circle \eqref{wcft:canonicalcircle} and thermal circle \eqref{wcft:thermalcircle} can be computed by exploiting modular covariance in the small $\beta$ limit~\cite{Detournay:2012pc,Castro:2015csg}%
\eq{
Z(\beta, \theta)= \bigg[  \frac{k \theta^2}{4{\beta}} -  \bigg( \frac{2 \pi i{\theta}}{{\beta}}  \bigg) \mP_0^{vac} - \frac{4\pi^2}{{\beta}} \mL_0^{vac} \bigg].
}
We then find that the vacuum expectation values of the zero-mode charges $\mL_0$ and $\mP_0$ are related to the thermodynamic potentials $\beta$ and $\theta$ by\footnote{For convenience, we have written $\mL_0$ and $\mP_0$ instead of  $ \langle \mL_0 \rangle_{\beta,\theta}$  and $ \langle \mP_0 \rangle_{\beta,\theta}$ where $ \langle \,\, \rangle_{\beta,\theta}$ denotes the thermal expectation value.} 
  \eqsp{
  \mL_0 &= -\frac{\p \log Z (\beta, \theta)}{\p \beta} = \frac{k}{4} \frac{\theta}{\beta^2} - \frac{2\pi i  \theta}{\beta^2} \mP_0^{vac} - \frac{4 \pi^2}{\beta^2} \mL_0^{vac}, \\
  \mP_0 &= \frac{\p \log Z (\beta, \theta)}{\p \theta} = \frac{k}{2} \frac{\theta}{\beta} - \frac{2\pi i }{\beta} \mP_0^{vac}. \label{wcft:thermalvevs}
  }
Using the values of $\mL_0^{vac}$ and $\mP_0^{vac}$ given in \eqref{wcft:vacuumcharges}, the equations above imply that
  \eq{
  \frac{\delta \beta}{\beta^3} = - \frac{3}{c} \frac{1}{\pi^2} \delta \bigg(\mL_0 - \frac{\mP_0^2}{k}\bigg), \qquad \delta \bigg(\frac{\theta - 2\pi\mu}{\beta} \bigg) = \frac{2}{k} \delta \mP_0,
  }
whereupon the variation of the entanglement entropy reads
  \eq{
  \delta S_{\mA} = - \frac{\mu k}{2} \delta l_y - \mu l_x  \delta \mP_0 - \frac{\beta^2}{4\pi^2}  \bigg(2 - \frac{2\pi l_x}{\beta} \coth\frac{\pi l_x}{\beta} \bigg) \bigg[ \delta \mL_0 - \bigg(\frac{\theta - 2\pi\mu}{\beta} \bigg) \delta \mP_0 \bigg] . \label{wcft:deltaS1}
  }

On the other hand, using  eq.~\eqref{wcft:wcftcharges}, we find that for zero-mode fluctuations, variations of the Virasoro and $U(1)$ currents are proportional to variations of the zero-mode charges,
 \eq{
  \delta T(x) = - \delta {\cal L}_0, \qquad \delta P(x) = - \delta {\cal P}_0. \label{wcft:zeromodefluctuations}
  }
Plugging \eqref{wcft:zeromodefluctuations} into \eqref{wcft:deltaH}, it is straightforward to verify that the variation of the modular Hamiltonian matches the variation of the entanglement entropy \eqref{wcft:deltaS1}, namely
\eq{  
\delta \langle{\cal H}_{mod}\rangle =   \delta S_{\mA} .
} 
 Consequently, the first law of entanglement entropy holds for zero-mode fluctuations.   
 

\subsubsection*{Local excitations} 

We now show that the first law of entanglement entropy also holds for variations of the entanglement entropy that result from the insertion of a local operator. Let us consider a state $\Ket{\psi}$ created by inserting a primary operator $\mathcal O_{\psi}$ with conformal weight $h_\psi$ and charge $q_\psi$ at the origin of the reference plane~\cite{Song:2017czq}. The expectation value of the currents on the reference plane can then be written as 
\eq{ 
T_\psi^{pl}(z) = z^{-2} h_\psi, \qquad P^{pl}_\psi(z) = i z^{-1} q_{\psi}. \label{wcft:hpsidef}
}
Using the transformation rules \eqref{wcft:Tprime} and \eqref{wcft:Pprime} under the warped conformal transformation \eqref{wcft:thermalcc}, we obtain the following expectation values of the currents on the state $\Ket{\psi}$  on the thermal cylinder 
\eq{
 T_\psi(x)  &= \frac{4\pi^2}{\beta^2} h_\psi + \frac{2\pi i}{\beta} \Big(\frac{\theta - 2\pi \mu}{\beta}\Big) q_{\psi} - \frac{\pi^2}{\beta^2} \frac{c}{6} - \frac{k}{4} \Big(\frac{\theta - 2\pi \mu}{\beta}\Big)^2, \label{wcft:Tpsi}\\
{P_\psi(x)} &= \frac{2\pi i }{\beta}q_\psi - \frac{k}{2} \Big(\frac{\theta - 2\pi \mu}{\beta}\Big). \label{wcft:Ppsi}
}
In particular, for the vacuum state where $ h_{vac} =  q_{vac} = 0$, eqs.~\eqref{wcft:Tpsi} and \eqref{wcft:Ppsi} correspond to the vacuum expectations values of the currents on the thermal cylinder. These equations also imply that the variation of the currents due to an infinitesimal change of the operator $\mathcal O_\psi$ satisfy
  \eq{
 \delta T_\psi(x) - \bigg(\frac{\theta - 2\pi\mu}{\beta} \bigg) \delta P_\psi(x) = \frac{4\pi^2}{\beta^2} \delta  h_{\psi},    \qquad  \delta P_\psi(x) =  \frac{2\pi i}{\beta} \delta  q_{\psi}.  \label{wcft:deltaT}
  }
As a result, the variation of the modular Hamiltonian \eqref{wcft:deltaH} evaluated on the state $\Ket{\psi}$ is given by
  \eq{
\delta\langle {\cal H}_{mod}\rangle =  - \frac{\mu k}{2} \delta l_y + \frac{2\pi i\mu l_x}{\beta} \delta q_{\psi} +  \bigg( 2 - \frac{2\pi l_x}{\beta} \coth \frac{\pi l_x}{\beta}\bigg) \delta h_{\psi},
  }
On the other hand, expanding the entanglement entropy of an excited state \eqref{wcft:psiee} to linear order in $\delta h_\psi$ and $\delta q_\psi$, we find an agreement with the variation of the modular Hamiltonian
  \eq{
  \delta S_{\cal A}^{\psi} = \delta \langle{\cal H}_{mod} \rangle,  
  }
which establishes the first law of entanglement entropy in holographic WCFTs that result from  the insertion of a local operator.


\subsection{Bulk modular flow and the first law} \label{wcftse:bulkmodularH}

In this section we consider the bulk side of the AdS$_3$/WCFT correspondence, write down the modular flow generator associated with a single interval on the boundary, and compute the variation of the corresponding gravitational charge. We will show that the latter matches the variation of the modular Hamiltonian on the boundary and use it to establish the first law from the bulk side of the correspondence.

\subsubsection{Bulk modular flow and holographic entanglement entropy}

In order to extend the modular flow generator~\eqref{wcft:kt3} from the boundary into the bulk we first need to describe the holographic dictionary relating the bulk and boundary theories. The modular flow generator~\eqref{wcft:kt3} was defined on a thermal state in a WCFT, the latter of which is dual to a BTZ black hole. In the Fefferman-Graham gauge the BTZ metric reads
  \eq{
  ds^2 = \frac{dr^2}{r^2} + \bigg( r^2 + \frac{ T_u^2 T_v^2}{r^2} \bigg) du dv + T_u^2 du^2 + T_v^2 dv^2,  \label{wcft:btzmetric}
  }
where $(u,v) \sim (u + 2\pi, v + 2\pi)$ and the endpoints of the interval $\mA$ are parametrized by
  \eq{
    \p \mA=\{ (u_-,v_-),(u_+,v_+)\}, \qquad l_u \equiv  u_+ - u_-, \quad l_v \equiv v_+ - v_-.   \label{wcft:interval3}
   }
 The parameters $T_u$ and $T_v$ in~\eqref{wcft:btzmetric} are related to the charges and thermodynamic potentials $\beta$ and  $\theta$ of a thermal state in the dual WCFT by the following holographic dictionary~\cite{Song:2016gtd}
  \eqga{
  \beta  = \frac{\pi}{T_u}, \qquad \theta = 2\pi\mu\bigg(\frac{T_v}{T_u} + 1\bigg),  \qquad   \mu =  \sqrt{-\frac{c}{6k}\,}, \\
   l_u = l_x, \qquad l_v = l_x + \frac{1}{2 \mu T_v} l_y, \label{wcft:dictionary}
  }
where the second line corresponds to eq.~\eqref{wcft:intervalmap} with $\mP_0 = k \mu T_v$.
  
We can extend the modular flow generator~\eqref{wcft:kt3} into the bulk by expressing it in terms of the local Killing vectors of the BTZ background. The Killing vectors of \eqref{wcft:btzmetric} compatible with the CSS boundary conditions are given by
  \eq{
   \!\! \!\! {\bar L}_0 = -\frac{\p_v}{2T_v}, \,\,\, \, L_0 = -\frac{\p_u}{2 T_u},\,\,\, \, L_{\pm} = -\frac{e^{\pm 2 T_u u}}{2T_u} \bigg [ \frac{r^4 + T_u^2 T_v^2}{r^4 - T_u^2 T_v^2} \partial_u -  \frac{2T_u^2 r^2}{r^4 - T_u^2 T_v^2}  \partial_v \mp T_u r \partial_r \bigg ],   \label{wcft:killing}
  }
where the normalization of the $L_0$, $L_{\pm}$ Killing vectors is chosen to satisfy the $SL(2,R) \times U(1)$ subalgebra of the asymptotic Virasoro-Kac-Moody algebra,
  \eq{
  [L_+, L_-] = 2L_0, \qquad [L_0,L_{\pm}] = \mp L_{\pm}, \qquad [\bar{L}_0, L_0] =  [\bar{L}_0, L_{\pm}] = 0.
  }
The linear combination of $SL(2,R) \times U(1)$ symmetry generators that asymptotes to the modular flow generator~\eqref{wcft:kt3} at the boundary is given by
  \eq{
  \xi & = 2\pi \bar{L}_0 + a_0 L_0 + a_{-1} L_+ + a_{1} L_{-} , \notag \\
  & =  2\pi \bar{L}_0 - 2 \pi \coth(T_u l_u) L_0 + \frac{\pi}{\sinh T_u l_u} \big[ e^{-T_u(u_+ + u_-)} L_+ + e^{T_u(u_+ + u_-)}  L_- \big], \label{wcft:ktbulk}
  }
where we used the holographic dictionary~\eqref{wcft:dictionary} and the $a_i$ coefficients can be obtained from \eqref{wcft:a_i} by a warped conformal transformation~\eqref{wcft:thermalcc}. 
 
The modular flow generator $\xi$ features a bifurcating Killing horizon
\eq{
N_\pm:\, r^2 = \mp T_u T_v \coth\big[ T_u (u - u_\pm) \big],
}
and it vanishes on the bifurcating surface $\gamma_{\xi}$, which corresponds to a line that extends along the $v$ direction and lies at a finite radial distance from the origin,
  \eq{
  \gamma_{\xi} : \, u = \frac{u_+ + u_-}{2}, \quad r = \sqrt{T_u T_v \coth T_u l_u/2}.\label{wcft:gammainf}
  }
The surface $\gamma_\xi$ is extremal but, in contrast to the standard RT/HRT surface in AdS/CFT, it is not homologous to the interval $\mA$ at the boundary. Finally, it is not difficult to check that the bulk modular flow generator~\eqref{wcft:ktbulk} satisfies the general properties described in Section~\ref{se:genrt}, namely eqs.~\eqref{normaleq1} and~\eqref{normaleq2}. 

\bigskip
\noindent{\bf Modular scrambling modes.} Let us now comment on the variations of the modular flow generator considered in Section~\ref{wcftse:modularalgebra}. In the bulk, the variation of the modular flow generator \eqref{wcft:ktbulk} is given by
  \eq{
  \delta_\pm \xi = \delta u_\pm \partial_{u_\pm} \Big( \sum_{i=-1}^1 a_i L_i + 2\pi \bar{L}_0 \Big) = \delta u_\pm \partial_{u_\pm} \xi. \label{wcft:deltaxi}
  }
In analogy with the flat$_3$/BMSFT case considered in Section~\ref{se:bulkflowandHEE}, the variations \eqref{wcft:deltaxi} satisfy a similar equation to \eqref{wcft:modularalgebra} with respect to the Lie bracket
\eq{
[\xi, \delta_\pm \xi ]=\pm 2\pi  \delta_\pm \xi, \label{wcft:bulkscrambling}
}
which is isomorphic to the algebra between a boost generator and a null translational generator of the effective Poincar\'{e} algebra in a small region of the spacetime \cite{deBoer:2019uem}. We also find that $\delta_+ \xi$ and $\delta_- \xi$ are two independent null vectors satisfying
\eq{
 \delta_+ \xi \cdot \delta_+\xi  & = \delta_- \xi \cdot \delta_-\xi  = 0, \qquad \delta_+ \xi \cdot \delta_-\xi \neq 0, \label{wcft:nullvector}
 }
and that  $\delta_\pm \xi$ lies on the light-sheet $N_\mp$, which follows from
 \eq{
 \delta_\pm \xi \cdot n_{N_\mp}&=0,
 }
 where $n_{N_\mp}$ is the vector normal to $N_\mp$. We thus find that $\delta_\pm \xi$  generates a null deformation of $\gamma_\xi$ along $N_{\mp}$. In other words, the mode $\delta_+ \xi$ generates a deformation of $N_+$ that does not affect $N_-$. At the boundary, this corresponds to a deformation of the boundary of the causal domain $\cal D$ located at $u = u_+$, while the other boundary at $u = u_-$ is kept fixed. Similar statements hold for $\delta_- \xi$. Hence, the behavior of $\delta_\pm \xi$ is compatible with the behavior of the modular scrambling modes $\delta_\pm \zeta$ discussed in Section~\ref{wcftse:modularalgebra}. Although the discussion here is valid for cases where $\xi$ is an exact Killing vector, it is straightforward to generalize these results to the more general cases considered in \cite{Apolo:2020bld}. In that case, eqs.~\eqref{wcft:bulkscrambling} and \eqref{wcft:nullvector} remain valid near the corresponding swing surfaces, so that the geometric interpretation of the modular scrambling modes still holds.

\bigskip

\noindent{\bf Holographic entanglement entropy.} Following the general discussion in Section~\ref{se:genrt}, we can construct a swing surface $\gamma_{\cal A}$ that is homologous to $\mA$ and is made up of the union of the null geodesics $\gamma_{\pm}$ tangent to the bulk modular flow plus the spacelike geodesic $\gamma$ connecting $\gamma_{-}$ to $\gamma_{+}$ along $\gamma_{\xi}$, namely
  \eq{
  \gamma_{\cal A} = \gamma_- \cup  \gamma \cup \gamma_+.
  }
 The null ropes $\gamma_{\pm}$ can be conveniently parametrized in terms of  the $u$ coordinate by   
  \eq{
  \!\!\!\!   \! \gamma_{\pm}: \, v &= v_\pm \pm  \dfrac{1}{2 T_v} \log \bigg \{ \dfrac{ \sinh \big[ \pm 2T_u (u_\pm - u) \big]} { \varepsilon_{} T_u }\bigg\} ,   \,\,\,\,\, r = \sqrt{T_u T_v \coth \big [ \pm T_u (u_\pm - u) \big]  },   \label{wcft:gammaplusminus}
  }
where $2 u \in [u_+ + u_-,\, 2u_+ - \varepsilon ]$ for $\gamma_+$  while $2u \in  [ 2u_- + \varepsilon , \,  u_+ + u_- ]$ for $\gamma_-$. In particular, the null geodesics $\gamma_{\pm}$ intersect the asymptotic boundary  at
  \eq{
(u,v,r) = \bigg( u_\pm \mp \frac{\varepsilon}{2}, \, v_\pm, \, \sqrt{\frac{2 T_v}{\varepsilon}} \,\bigg).
  }
 On the other hand, the bench $\gamma$ of the swing surface is a line that can be parametrized by
\eq{
  \gamma&: \, u = \frac{u_+ + u_-}{2}, \quad v = \frac{v_+ + v_-}{2} + \frac{\lambda}{2 T_v} \log \bigg( \frac{ \sinh T_u l_u} { \varepsilon_{} T_u }\bigg) , \quad  r = \sqrt{T_u T_v \coth \frac{T_u l_u }{2}  }, \label{wcft:gammaextremal}
  }
where $\lambda \in [-1,\,1]$. 
Using the holographic dictionary~\eqref{wcft:dictionary}, it is not difficult to verify that the area of the swing surface in Planck units equals the entanglement entropy computed at the boundary~\eqref{wcft:thermalee}, namely
  \eq{
  S_{\cal A} (\beta, \theta) = \frac{\mathrm{Area}(\gamma_{\cal A})}{4 G}. 
  }
  %


\subsubsection{The first law in AdS$_3$/WCFT} \label{wcftsuse:bulkmodH}

In this section we compute the variation of the gravitational charge associated with the bulk modular flow generator \eqref{wcft:ktbulk} in Einstein gravity. We will show that this charge matches the variation of the modular Hamiltonian given in \eqref{wcft:deltaH}. Then, using the fact that the swing surface $\gamma_{\cal A}$ is homologous to the boundary interval $\mA$, we provide a holographic derivation of the first law of entanglement entropy in AdS$_3$/WCFT.

In the covariant formulation of gravitational charges, the variation of the gravitational charge associated with the bulk modular flow generator $\xi$ is defined by
  \eq{
  \delta {\cal Q}_\xi^{\cal A}[g] =  \int_{\cal A} {\bm \chi}_{\xi} [g,\delta g],
  }
where $ {\bm \chi}_{\xi} [g,\delta g]$ is a one-form given in \eqref{chidef} and the variation $\delta g$ of the background metric $g$ must be compatible with the boundary conditions. For pure Einstein gravity with CSS boundary conditions, the phase space in the Fefferman-Graham gauge is given by \eqref{wcft:cssfg} and we have
  \eqsp{
  \delta g_{\mu\nu} dx^{\mu} dx^{\nu} = & r^2 \delta  J'(u) du^2 + \delta  L(u) du^2  + 2 T_v^2 \delta  J'(u)  du dv +  \delta T_v^2 dv^2  \\
  & \hspace{1.5cm} + \frac{1}{r^2} \Big[ T_u^2 T_v^2 \delta J'(u) du^2 +  (T_v^2 \delta L(u) + T_u^2 \delta T_v^2 \big) du dv \Big],
  }
where $\delta T_v$ is independent of the coordinates as required by the CSS boundary conditions~\eqref{wcft:cssbc}. The one-form ${\bm \chi}_{\xi} [g, \delta g]$ is independent of the radial coordinate and the corresponding gravitational charge is given by
  \eq{
  \delta {\cal Q}_\xi^{\cal A} [g] = \frac{c}{6} \int_{\cal A} \bigg\{ \bigg [ \frac{1}{2 T_u}\bigg[ \frac{ \cosh l_u T_u - \cosh T_u (2u -u_+ -u_-)}{\sinh l_u T_u} \bigg] \delta L (u)  +   T_v \delta  J'(u) \bigg ] du    +  \delta T_v dv \bigg\}. \label{wcft:bulkH}
  }
We now have the necessary ingredients to establish the first law of entanglement entropy in AdS$_3$/WCFT. We first show that \eqref{wcft:bulkH} matches the variation of the modular Hamiltonian given in~\eqref{wcft:deltaH}. In order to see this we can use the holographic dictionary between the bulk and boundary currents \eqref{wcft:currentdictionary} to relate the bulk and boundary fluctuations 
  \eq{
  \delta T(x) &= -\frac{c}{6} \big [ \delta L(u) - \delta T_v^2 - 2 T_v^2 \delta J'(u) \big],\qquad \delta P(x)= - \mu k \big [\delta T_v + T_v \delta  J'(u) \big]. \label{wcft:fluctuations}
  }
One of the subtleties of the AdS$_3$/WCFT correspondence is that the map between the bulk and boundary coordinates \eqref{wcft:map} depends on the state. Since $l_u$ and $l_v$ are held fixed in the derivation of \eqref{wcft:bulkH}, we find that
   \eq{
   \delta l_u = \delta l_v = 0 \quad \implies \quad \delta l_u =0, \quad \delta l_y = 2\mu (l_v - l_u) \delta T_v. \label{wcft:lengthvariations}
   }
Then, using eqs.~\eqref{wcft:fluctuations} and~\eqref{wcft:lengthvariations}, we find that the gravitational charge in the bulk matches the modular Hamiltonian at the boundary,
  \eq{
  \delta {\cal Q}_\xi^{\cal A} [g] = \delta \langle{{\cal H}_{mod}}\rangle, \label{wcft:matchH}
 }
thereby explicitly verifying the general discussion of Section~\ref{se:generalfirstlaw} in the context of the AdS$_3$/WCFT correspondence.
 
Since the swing surface $\gamma_\mA$ is homologous to the boundary interval $\mA$, Stoke's theorem guarantees that the gravitational charges evaluated on $\gamma_\mA$ and $\mA$ agree with one another. In particular, for constant fluctuations of the zero-modes where $\delta L (u) = \delta T_u^2$ and $\delta J (u) = 0$ we have
  \eq{
\delta {\cal Q}_\xi^{\gamma_{\cal A}} [g] =\delta {\cal Q}_\xi^{\cal A}[g] &= \delta \bigg\{  \frac{c}{6} l_v T_v + \frac{c}{6} \log \bigg( \frac{1}{T_u \varepsilon} \sinh l_u T_u \bigg) \bigg \}=\delta S_{\cal A}  (\beta, \theta), \label{wcft:varwaldentropy}
  }
  where in the last equality we have used the holographic dictionary \eqref{wcft:dictionary} and the expression for the holographic entanglement entropy  \eqref{wcft:thermalee}.
This provides an explicit check for the assumption \eqref{s2:dsQA} which states that the entanglement entropy can be calculated using the gravitational charge associated with the bulk modular flow generator $\xi$ evaluated on the swing surface $\gamma_{\cal A}$. Finally, combining eqs.~\eqref{wcft:matchH} and \eqref{wcft:varwaldentropy}, we establish the first law of entanglement entropy in AdS$_3$/WCFT,
 \eq{
  \delta S_{\cal A}= \delta {\cal Q}_\xi^{\gamma_{\mA}}[g]=\delta {\cal Q}_\xi^{\mA}[g]= \delta \langle{\cal H}_{mod} \rangle.
  }
as described in general in Section~\ref{se:generalfirstlaw}.
 

\subsection{Warped AdS} \label{se:wads}

We conclude by noting that the results derived in this section can be generalized to warped AdS$_3$ (WAdS$_3$) spacetimes whose metric is given by~\cite{Detournay:2012dz}
\eq{
\!\!\! ds^2 = \frac{(1+\lambda^2 T_v^2)dr^2}{r^2} + \bigg( r^2 + \frac{ T_u^2 T_v^2}{r^2} \bigg) du dv + \bigg[ T_u^2 -\frac{\lambda^2 \left(r^4-T_u^2 T_v^2\right)^2}{4 r^4} \bigg] du^2 + T_v^2 dv^2. \label{wcft:wadsmetric}
}
Indeed, provided we use the bulk modular flow generator~\eqref{wcft:ktbulk}, we find that  the same conditions in eqs.~\eqref{normaleq1} --~\eqref{surfacegravity} are satisfeid in the WAdS$_3$ background~\eqref{wcft:wadsmetric}. Consequently, the geometric picture for holographic entanglement entropy in WAdS$_3$ is the same as in AdS$_3$ and the swing surface is described by the same set of null ropes~\eqref{wcft:gammaplusminus} and the same bench~\eqref{wcft:gammaextremal} found earlier. 

As a simple example, we can consider the so-called $S$-dual dipole theory \cite{Detournay:2012dz},
\eq{
\!\! S[g,U,A]  = \! \! \int \! d^3 x \sqrt{|g|}  \bigg[ R - 4(\partial U)^2+\frac{2}{\ell^2}e^{4U}\Big(2-e^{4U}\Big)-\frac{1}{\ell}\epsilon^{\mu\nu\rho}A_{\mu}F_{\nu\rho} -\frac{4}{\ell^2}A_{\mu}A^{\mu} \bigg],
}
where we have set $16\pi G = 1$ for convenience, $U$ is a scalar, and $F_{\mu\nu}$ is the field strength of the abelian vector field $A_{\mu}$. This theory admits the above WAdS$_3$ metric~\eqref{wcft:wadsmetric} as a solution together with the following values of the scalar and vector fields
  \eq{
e^{-4U}&=1+\lambda^2 T_v^2,\qquad A=\frac{\lambda \ell}{1+\lambda^2 T_v^2}\bigg(T_v^2 dv + \frac{r^4+T_u^2 T_v^2}{2r^2} du\bigg).
}
When restricted to the sector of the phase space without bulk propagating modes, i.e.~when $U$ is constant, the validity of the holographic entropy proposal discussed at the geometric level above, is enough to guarantee the validity of the expressions for the entanglement entropy, modular Hamiltonian, and the first law found in previous subsections. In order to show this, we can use the map described in~\cite{Compere:2014bia} between the background fields of the $S$-dual dipole theory~$(g,U,A)$ and an auxiliary AdS$_3$ metric~($\hat{g}$) in Einstein gravity, namely
\eq{
	g_{\mu\nu}=e^{-4U}\big(\hat{g}_{\mu\nu}-A_\mu A_\nu \big) \label{wcft:wadstoads}.
}
Such a map induces an isomorphism between the phase spaces of pure Einstein gravity and the sector without bulk propagating degrees of freedom in the $S$-dual dipole theory. In particular, this isomorphism preserves the symplectic structure and leads to the same gravitational charges as in pure Einstein gravity. Furthermore, under this isomorphism the WAdS$_3$ background~\eqref{wcft:wadsmetric} is mapped to an auxiliary AdS$_3$ metric $\hat{g}_{\mu\nu}$ which is exactly the BTZ solution~\eqref{wcft:btzmetric} with the same values of the $T_u$ and $T_v$ parameters. Thus, when evaluated on the same swing surface $\gamma_{\cal A}$, the gravitational charges yield the same results as in AdS$_3$, namely, the same entanglement entropy, modular Hamiltonian, and the first law.


\bigskip

\section*{Acknowledgments}
\noindent We are grateful to Pankaj Chaturvedi, Bartek Czech, Stephane Detournay, Daniel Harlow, Juan Maldacena, Prahar Mitra, Max Riegler, and Herman Verlinde for helpful discussions. LA and WS thank the Kavli Institute of Theoretical Physics for hospitality and for providing a stimulating environment during the program ``Gravitational Holography''. LA also thanks the Institute for Advanced Study for their kind hospitality. The work of LA, WS, and YZ was supported by the National Thousand-Young-Talents Program of China, NFSC Grant No.~11735001, and Beijing National Science Foundation No.~Z180003. LA and WS were supported in part by the National Science Foundation under Grant No.~NSF PHY-1748958. The work of LA was also supported by the International Postdoc Program at Tsinghua University and NFSC Grant No.~11950410499. HJ is supported by the Swiss National Science Foundation.


\bibliographystyle{JHEP} 
\bibliography{BMS_WCFT,refs}

\end{document}